\documentclass[twocolumn,nofootinbib]{revtex4-2}

\usepackage{amsmath}
\usepackage{amsfonts}
\usepackage{amssymb}
\usepackage{graphicx}
\usepackage{scrextend}
\usepackage{multirow}
\usepackage{color}
\usepackage{enumitem}
\usepackage{multibib}

\usepackage[hidelinks,colorlinks=true,allcolors=blue,urlcolor=blue]{hyperref}
\usepackage{verbatim}
\usepackage{lineno}

\newcommand{\D}{\mathrm{d}}

\newcommand{\eq}{Eq.}

\newcommand{\fig}{Fig.}

\newcommand{\figheight}{0.3} 

\usepackage{simplemargins}
\settopmargin{1.8cm}
\setbottommargin{2.4cm}

\AtBeginDocument{%
\setlength{\abovedisplayskip}{5pt}
\setlength{\belowdisplayskip}{5pt}

\setlength{\belowcaptionskip}{-6pt}
}

\begin{document}


\title{A stop to field line misconceptions}

\author{Petar \v{Z}ugec}
\affiliation{University of Zagreb Faculty of Science, Department of Physics, Zagreb, Croatia}

\author{Ivica Fri\v{s}\v{c}i\'{c}} \affiliation{University of Zagreb Faculty of Science, Department of Physics, Zagreb, Croatia}
\author{Mihael Makek} \affiliation{University of Zagreb Faculty of Science, Department of Physics, Zagreb, Croatia}
\author{Eric Andreas Vivoda} \affiliation{University of Zagreb Faculty of Science, Department of Physics, Zagreb, Croatia}


\begin{abstract}

For two hundred years---ever since Faraday's first conception of the field lines---hearts and minds of students have been pervaded by erroneous and unsubstantiated claims about these well defined mathematical objects. The most prominent misconceptions include: (1)~a notion of the field lines as of `lines of force'; (2)~a notion that the field lines coincide with particle trajectories; (3)~a notion that a field line density measures a magnitude of a vector field; (4)~a notion that the field lines of a divergence-free field always form closed loops. Even the modern day literature systematically perpetuates some of these claims. We compile here the \textit{trivial} counterexamples to these claims, providing physics instructors with an efficient and effective way of dispelling these misconceptions.

\end{abstract}

\maketitle

\section{Introduction}
\label{intro}

Ever since Michael Faraday introduced the idea of field lines in 1831~\cite{faraday,lines_force}---under the name and context of \textit{lines of force}---they have been a constant subject of academic discussions. Initially they were thought of as true physical entities, worthy of earnest scientific consideration, until they gave way to a more fundamental notion of a \textit{vector field}. In that, Faraday's `lines of force' were an important stepping stone toward the formation and acceptance of a field concept, championed by the success of Maxwell's work on electrodynamics~\cite{field_history}. As such, their importance in a historical context---and, indeed, in a context of a logical development of the idea of field---should not be downplayed in any way. However, in two hundred years our understanding of fields and field lines has evolved significantly beyond Faraday's initial conceptions. 

Despite that, various misconceptions about the nature of field lines still persist both among the students and throughout the literature, educational or otherwise. By definition, a field line is a curve that is at each point along its length tangential to a given vector field. Throughout literature they are almost always defined \textit{only} by this verbal description, without providing any further mathematical formalization, i.e. any type of operational equation. Though this description is semantically sufficient---in a sense that the general differential field line equation [\eq~(\ref{master}) from Appendix~\ref{appendix_A}] may be extracted from it by a careful deliberation---the form of this equation is not immediately obvious. Without such operational tool the field lines remain ill-defined in all but name, which is a major source of the associated misconceptions.

In the last few decades an extensive body of work has been produced on students' understanding of fields and field lines~\cite{lines_force,trajectory_0,trajectory_1,trajectory_3,trajectory_4,students_0,students_1,students_2,students_3}. In many ways their grasp on both concepts was found to be lacking. For example, students often confuse the field lines with the real physical entities, assigning them the matter-based properties and treating them as tubes for transmitting the forces, electrical action, electric charges, etc.~\cite{trajectory_0,lines_force,trajectory_3}. Specially notable is the confusion with the particle trajectories~\cite{trajectory_0,trajectory_1,lines_force,trajectory_3,trajectory_4}---one of the major misconceptions addressed in this work. Students also seem to struggle with the conversion between the field lines and the associated vector fields, especially when the field superpositions are involved~\cite{students_0,students_1,students_2,students_3}. Other difficulties include a failure to recognize that the field lines may not intersect, or even that the vector field effects are also present in the gaps between the drawn field lines. 


A major source of these misconceptions is the fact that the field lines are not a \textit{faithful representation} of vector fields. That they do not represent \textit{all} the properties of a given vector field is of lesser importance. Of greater impact is that they bring their own ``excess baggage''---the properties that do \textit{not} belong to the vector field itself---thus leading to the confusion by representation~\cite{trajectory_0}. That the educational literature is not consistently `up to par' on the concepts of vector fields and of field lines---often suffering from some weaknesses in expositions~\cite{field_history} or from outright errors~\cite{electric_1,magnetic_1,magnetic_3,magnetic_4}---certainly does not help in resolving the issue. In fact, even some of the dedicated studies on students' struggles with fields and field lines seem to perpetuate certain misconceptions on the subject.

The main part of this work addresses the most widespread misconceptions about the field lines. These include false notions that: (1)~field lines have something to do with a force; (2)~field lines correspond to the particle trajectories within a vector field; (3)~a density of field lines reflects a magnitude of a vector field; (4)~the field lines of a divergence-free vector field (such as a magnetic field) always form closed loops. \textit{One could take almost any nontrivial example of a vector field and find it to be a counterexample to these claims.} However, a detailed calculation of field lines from such fields obviously requires some focused effort. Hardly anyone, save those utterly fascinated by a given subject, is interested in these kinds of highly technical proofs and demonstrations. Students in particular will not be captivated by the (counter)examples they can not mentally reproduce at a moment's notice. For this reason we will give here \textit{trivial counterexamples} to these claims. They can be easily internalized by the students and always be `held at the ready' by the instructors.

Considering such fuss about the field lines, it is only natural to ask if they are worth all the effort of introducing them. A convincing argument can certainly be made for their pedagogical merits~\cite{students_1}. They provide a visually attractive means of representing the vectors fields, thus partially solving a problem of the abstract nature of a field concept. In other words, they serve to lessen the intellectual shock when transitioning from classical mechanics to conventionally structured courses on classical electrodynamics. On the other hand, some authors argue that the field lines should be altogether abandoned from pedagogical practices and that the introductory courses on electromagnetism should be restructured~\cite{against_fl}. We can appreciate both sides of the argument.

The fact that the field lines \textit{were} an integral part of a historical development of a field concept strongly suggests that they are indeed a natural part of this creative process. This in itself is a sound argument for using them in a pedagogical, deductive discovery of real physical fields. For this reason we have a feeling that field lines are not going anywhere, anytime soon. In light of this, one needs to take preemptive measures against developing the wrong ideas about them; a sort of intellectual vaccination against false notions abounding in the intellectual ecosystem. Aside from discussing what field lines \textit{are}, it should also be made perfectly clear what they \textit{are not}. This is especially important when one considers a plethora of false claims not only transmitted by word of mouth but also firmly entrenched in literature. Ever since the 1950s---apparently starting with~\cite{redundant}---many authors have recognized and attempted to counter these issues, focusing mostly on the misconceptions and false claims about the electric~\cite{redundant,electric_1,electric_2,electric_3,electric_magnetic} and magnetic~\cite{redundant,electric_magnetic,magnetic_1,magnetic_2,magnetic_3,magnetic_4,ajp_new} field lines. Backed up by decades of these expositions---most of them focused on specific issues; some of them quite technical, some going into painful detail---we comprehensively address the most prominent file line misconceptions. For each of them we provide a \textit{simple} counterargument or a \textit{trivial} counterexample, thus removing a possibility of any further confusion about these issues. In the famous words by John S. Bell~\cite{bell}: `like all authors of noncommissioned reviews, [we think that we] can restate the position with such clarity and simplicity that all previous discussions will be eclipsed'.

\vspace*{-3mm}

\section{Field lines as the lines of force}
\label{force}

\vspace*{-1mm}

Of all misconceptions that we will address, the one about the field lines having something to do with a physical force seems to be the least explicit, though it certainly does seem to be present. It is clear from a historical name `the lines of force' that this idea was tacitly encouraged, even enforced in the past\footnote{
Among the references cited in this work, a consistent use of the term `the lines of force' may be found at least up to the 1950s~\cite{magnetic_1,redundant}. Equivalent term still persists in some languages to this day, as is the case with the authors' mother tongue.
}. Even some contemporary books still use it, treating is as an equivalent, if not a preferred term~\cite{book_jones}. Yet, even with a more appropriate name `the field lines', the idea about their connection with a force is \textit{implicitly} present within the next misconception on our list, regarding the particle trajectories (Section~\ref{trajectories}). By disentangling a concept of field lines from a concept of force, one goes a long way toward resolving the misconception about the particle trajectories. Therefore, this notion about the field lines and the force needs to be addressed as soon as possible in pedagogical expositions.

In fact, any discussion about the field lines should start by pointing out a well recognized fact that they are in no manner physical. They are artefacts of imagination, purely mathematical objects---deserving of intellectual pursuit in their own right---but still just the physicists' ephemeral `flights of fancy'. They do not correspond to nor do they describe, explain or give rise to any observable phenomena. Nor are they themselves subject to any physical law. If they were in any manner physical, one should be able to formulate a physically meaningful concept of their motion. There is no unique way of achieving this, meaning that any such concept is necessarily arbitrary, basically as good as any other. In that, the intuitively appealing and otherwise reasonable definitions of the field line motion may lead to a field line ``propagation'' faster than light~\cite{lines_motion}. This suggests that the field lines are not only \textit{not subject} to any physical law, but that they \textit{can not} be.

It should also be well understood that the field lines are \textit{informationally redundant}~\cite{redundant}. A vector field \textit{completely} describes a given vectorial quantity, assigning to each point in space a magnitude and a direction. There is no room left for any other degree of freedom; none are left unaccounted. Thus the field lines do not---and can not---supplement a vector field by any objective information not already contained in a field itself; they only bring to light our own wishful thinking about it. As Slepian poetically says in describing Maxwell's endorsement of electric and magnetic fields in place of Faraday's `lines of force'~\cite{redundant}:\\ 

\begin{addmargin}[1em]{1em}
`Gone from the equations were the continuous lines of force [...]. There remained only the vital essence, the vector electric and magnetic fields, each giving at each point of space a magnitude and a direction and that is all.'\\
\end{addmargin}

A supposed connection between the field lines and a physical force is easily dismissed on a purely logical level. However, attempting to address it in the first place is not as trivial. This is because students sometimes seem to be wholly unaware that they hold any presumption about the subject at all, making this misconception less explicit than the other ones. However, their `subscription' to this idea---or the lack thereof---is clearly exposed when they are prompted about it, either directly or indirectly (through a supposed connection with the particle trajectories). We propose here a flow of discussion to be had with students in order to convincingly dispel this idea. It is designed as a series of counterarguments to the successive attempts at preserving the ingrained preconceptions about the `lines of force'. As such, the discussion progresses from more abstract arguments to the more specific ones, since we expect particular students to be convinced at different points in a discussion, if they be open to convincing at all.

\pagebreak

(1) As their present-day name suggests, the field lines are attributable to an arbitrary type of vector field, rather than being exclusive to the fields of force. Therefore, we may imagine an arbitrary vector field, as a pure mathematical construction without a physical reality. Since this field also has its own field lines\footnote{
We always assume well-behaved fields, such that the formal conditions for the existence of field lines are satisfied. We address a technical procedure for calculating the field lines in Appendix~\ref{appendix_A}.
}, they can not have any exclusive relation to a physical force.

From a purely logical standpoint, this argument is conclusive. However, it is so abstract and general (that being its objective strength) that students may find it unrelatable and unconvincing (thus subjectively misinterpreting it as weak). At this point instructors may expect counterclaims such as: `Okay, but what if we only consider physical fields'? The implied assumption here is that every physical quantity constituting a field always has a corresponding field of force, and that this field of force is everywhere collinear to a field in question. This idea stems from imagining only electric and gravitational fields, since they are most often discussed and they do indeed satisfy this associated-force requirement. The following counterarguments serve to refute this assertion.

(2) Imagine that there are several such fields present simultaneously and that they are not collinear at all points. For example, we have both a gravitational and an electric field of diverging directions. Since these are different physical quantities, their fields can not be combined into a single one. They necessarily stay separated and each one has its own field lines, different from those of the other field. However, they both cause their own fields of force which \textit{do} combine into a single physical field of force. This total field of force is no longer collinear with either of the two fields. Therefore, it has its own set of field lines, which coincide neither with the gravitational nor the electric field lines.

Unconvinced students might attempt to circumvent this argument by counterclaims of the type: `Okay, but what if we consider each field (i.e. its associated field of force) separately, \textit{as if} it were the only one present, and then talk about \textit{its own} lines of force'? We certainly could make this kind of mental acrobatics, ignoring a physical reality of the situation. But it should be pointed out that this `escape into the unphysical' (considering the unphysical fields of `partial' force, while ignoring the only physical field of force present---that of the total force) is \textit{inconsistent} with an earlier insistence on the physical reality of the considered fields, which slowly but surely leads us into a realm of intellectual dishonesty. In any case, at a price of inconsistent principles (an oxymoron, if ever there was one), this would certainly solve the problem for gravitational, electric or any similar field, seemingly allowing a preservation of one's fixed idea about the `lines of force'. However...

(3) There do exist physical vector fields that do have an associated force which is \textit{not} collinear to a field in question. A particular example is a magnetic field, giving rise to the Lorentz force which---if present at all---is at each point perpendicular to a magnetic field. In fact, the Lorentz force does not even constitute a field, since---by depending on a charged particle velocity---it can not be \textit{uniquely} assigned to any given point in space (except for trivial points where magnetic field vanishes). Thus, there \textit{is} a physical force associated with a magnetic field, but its `lines of force' do not exist since it is not a field. Even if we attempted to forcefully calculate them by artificially fixing a particle velocity in a familiar expression for the Lorentz force (\mbox{$\mathbf{F}=q\mathbf{v}\times\mathbf{B}$}), these lines of force would not correspond to the field lines of a magnetic field (which \textit{do} exist as well-defined mathematical objects). Any presumed universal connection between the field lines and the `lines of force' is thus severed.

Instructors should not be overly surprised by possible attempts at further (incoherent) counterclaims like: `Okay, but we could always \textit{mentally} equate a given vector field with a field of force, so that \textit{if} such associated field of force existed, we could treat the field lines of a field in question \textit{as if} they were the lines of force'. At this point it becomes clear that a student does not want and, thus, can not be convinced. One can only point out---as a last appeal to reason---that this `argument': (a)~is completely at odds with any earlier appeal to a physical reality; (b)~is self-contradictory since by entirely unphysical reasoning it insists on a connection with a physical reality (by insisting on a field of a physical force); (c)~comprises multiple other logical fallacies, such as \textit{special pleading} (pleading a physical reality for a purely mathematical object) and \textit{equivocation} (pronouncing any given field as a field of force). Adopting this kind of `reasoning', any desired claim can be proved or disproved (or both!), making it altogether worthless.

\section{Field lines as trajectories}
\label{trajectories}

Thankfully, a misconception about the field lines coinciding with particle trajectories does not seem to be perpetuated in modern literature. However, extensive studies show that it \textit{is} widespread among students~\cite{lines_force,trajectory_0,trajectory_1,trajectory_3,trajectory_4}. So much so that some sources specifically caution against it, and appropriately so~\cite{book_serway,book_young}. An explicit mention of it may also be found throughout past studies~\cite{electric_1,electric_2}, confirming that it is indeed an issue worth addressing. In fact, it is interesting to note how Freeman~\cite{electric_2} puts it, evidently being surprised by the fact himself: `Another discovery that is quite instructive is the fact that a free charge will not in general move along a field line'.

A discussion from Section~\ref{force} on a general disparity between the field lines and a physical force immediately severs any preconceived universal connection between the field lines and physical trajectories. Since field lines have no intrinsic connection to a force---not being in any way physical---they can not possibly always correspond to the force-affected physical trajectories, except by serendipitous chance.

\pagebreak

In that, it is not that the field lines \textit{never} coincide with the particle trajectories. Rather, they \textit{not always do}. In fact, given an arbitrary vector field---even the field of force---they \textit{almost never} do. There are special cases---such as a free fall from rest in a homogenous gravitational field---when the trajectory does indeed correspond to one of the gravitational field lines. Without such cases---and trivial ones at that!---the general misconception would hardly exist in the first place. The problem appears when these exceptional cases are \textit{all} that is ever considered. The fluid streamlines, for example, constitute an entire \textit{exceptional class} of field lines in this regard. Each steady-flow streamline does indeed correspond to a fluid-particle trajectory, but only by virtue of streamlines being the field lines of a very specific vector field: the \textit{velocity field}. Thus, a fluid-particle displacement \mbox{$\D\boldsymbol{\ell}=\mathbf{v}\D t$} alongside a steady-flow streamline is guaranteed by definition\footnote{
In the context of the differential field line equation~(\ref{master}) from Appendix~\ref{appendix_A}: $\D \mathbf{f}=\hat{\mathbf{v}}\D\ell$ for a given streamline~$\mathbf{f}$, so that \mbox{$\D\boldsymbol{\ell}=\D\mathbf{f}$} for a physical displacement~$\D\boldsymbol{\ell}$, with \mbox{$\D\ell=|\mathbf{v}|\D t$}.
}.

Of course, if the previous general argument---however correct and conclusive---was convincing and appealing to students, no misconception about the issue would exist in the first place. Therefore, we should aim to be more specific and persuasive. The next level of argumentation consists in pointing out that the field lines depend only on a vector field itself and the initial point selected for tracing out a specific line. On the other hand, particle trajectories require \textit{additional} initial conditions: not only the initial position but also the initial velocity! Imagine a homogeneous gravitational field (since by rejecting the previous argument a student still imagines a field of force, or at least a field that has an associated field of force). All field lines are mutually parallel straight lines. Now set a body in motion with initial velocity having a component perpendicular to a field. In other words, throw a ball in any direction other than a field direction. What is its trajectory and how does it relate to the field lines\footnote{You might also take advantage of the familiar elliptic planetary orbits within a central gravitational field of the nearest star, and of such field's field lines.}? 


At this point instructors should fully expect some more special pleading. Further `hidden variable' conditions that students might attempt are rather obvious: maybe field lines coincide with particle trajectories when the particle is let from rest (all the while thinking in terms of the `lines of force', having failed in disabusing that notion). Or maybe the trajectories coincide with the field lines for vanishingly `small' particles, i.e. in a limit of a vanishing mass, charge, etc. One can take any nontrivial example of a field of force and show by a direct calculation---just as Freeman did~\cite{electric_2}---that this is false. However, this requires some effort, making it a forever unpopular venue of persuasion. For this reason one needs a simple, specific counterexample that will illustrate a general untenability of these notions; a counterexample that is conclusive without any calculations.

\pagebreak

\begin{figure}[t!]
\centering
\includegraphics[height=\figheight\textwidth,keepaspectratio]{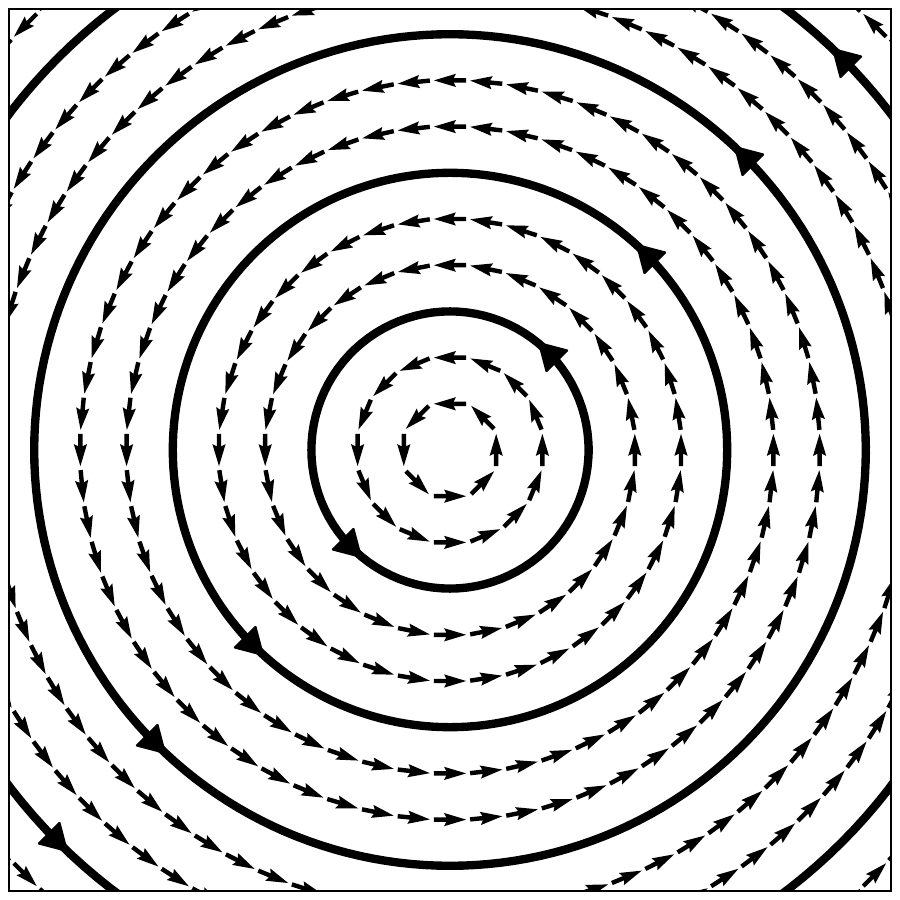}
\caption{First counterexample to a notion that field lines (solid directed lines) might coincide with particle trajectories. In this example only a field direction is relevant; it matters not if the arrow lengths represent a field magnitude or not.}
\label{fig1}
\end{figure}

Such counterexample is a `circular' field of force $\mathbf{F}(\mathbf{r})$, i.e. a field featuring only the azimuthal~($\hat{\boldsymbol{\varphi}}$) component:
\begin{equation}
\mathbf{F}(\mathbf{r})=F(\mathbf{r})\hat{\boldsymbol{\varphi}}.
\label{circular}
\end{equation}
In general, a field magnitude $F(\mathbf{r})$ may vary from point to point. Figure~\ref{fig1} shows such field, both by a vector plot (isolated arrows indicating a field direction at given points) and by several circular field lines (directed solid lines). If arrow lengths are taken as indicative of a field magnitude, and not only of a field direction, then the figure shows a field of a uniform magnitude: \mbox{$F(\mathbf{r})=\text{const.}$} The counterargument to a notion that the `lines of force' might coincide with particle trajectories is simple. \textit{If these circular field lines were trajectories, a force would have to have a centripetal component. However, the force imagined here is purely tangential to the supposed trajectories.} 

\begin{figure*}[t!]
\centering
\includegraphics[height=\figheight\textwidth,keepaspectratio]{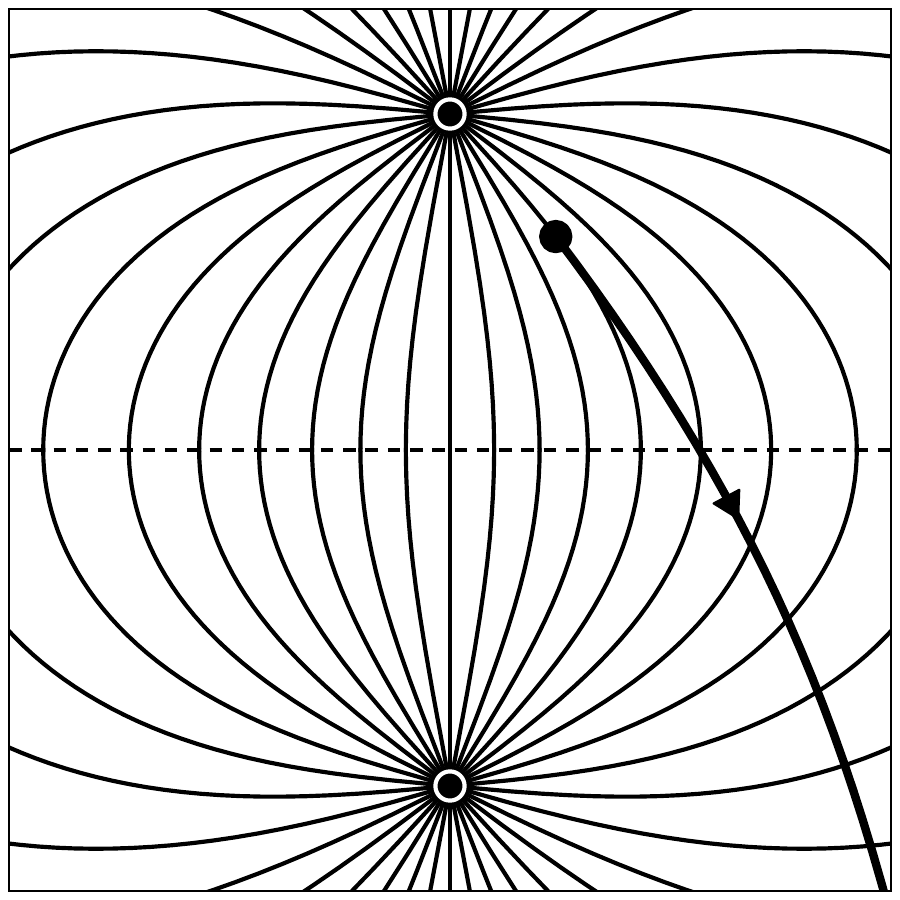}
\includegraphics[height=\figheight\textwidth,keepaspectratio]{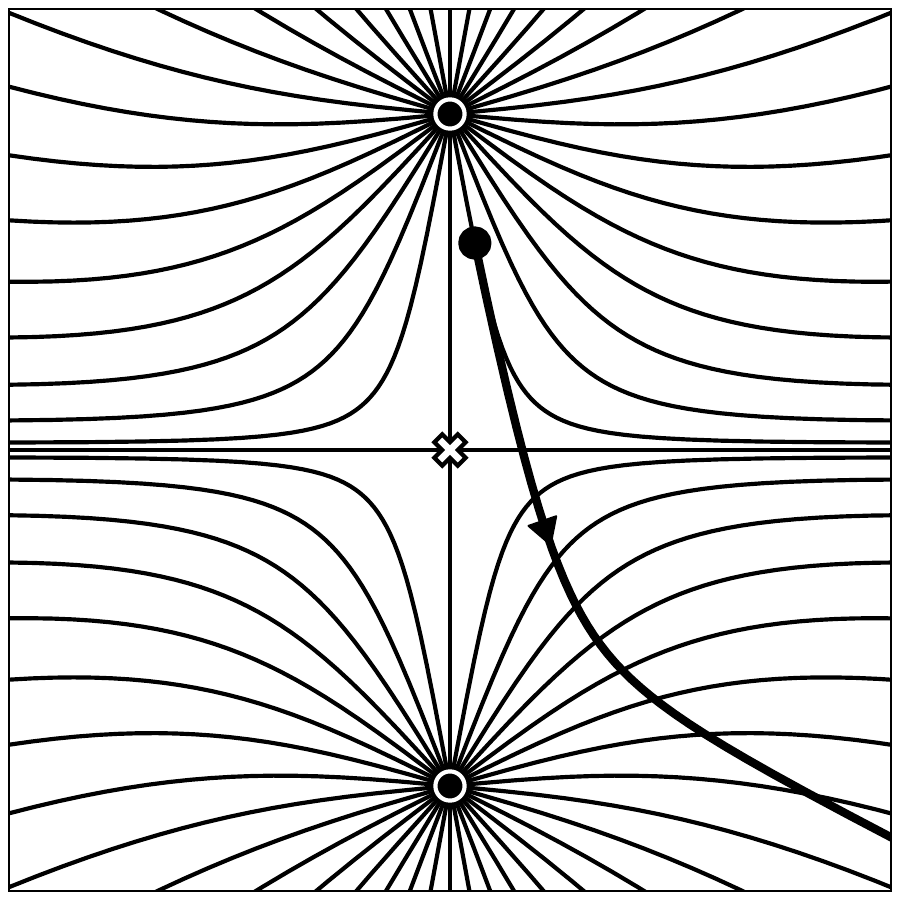}
\caption{Second counterexample to a notion that field lines might coincide with particle trajectories. Left panel: `attractive dipole' (\mbox{$q_1=-q_2$}). Right panel: `repulsive dipole'  (\mbox{$q_1=q_2$}). A test charged particle trajectory is shown by a thick directed line. The field lines emanating from mirrored charges $q_1$ and $q_2$ correspond to an electric field of a `dipole' configuration alone.} 
\label{fig2}
\end{figure*}

Though this is yet another conclusive proof, the instructors should be ready for further rebuttals of the type: `But this circular field is not (and can not be) a field of gravitational force nor of a familiar electrostatic force, caused by a static configuration of charges'. True, since both types of field are rationalness: \mbox{$\boldsymbol{\nabla}\times \mathbf{F}=\mathbf{0}$}. But notice that the counterargument extends well beyond circular fields. For \textit{any} curvilinear trajectory a force has to have a centripetal component. In other words, a curvilinear trajectory \textit{can not possibly} be tangential to a force, while the `lines of force' \textit{by definition are} tangential to it. Therefore, \textit{any} curvilinear `line of force' is a direct counterexample to a particle trajectory notion: \textit{no} curvilinear trajectory, in \textit{any} field of force---including gravitational or electrostatic one---can coincide with the `lines of force', regardless of the particle parameters (their `smallness'), regardless of the initial conditions (initial rest) and regardless of any other conceivable circumstance.

Students can always express a dissatisfaction due to the abstractness of the argument. All-encompassing generality is hard to visualize and internalize. It is difficult to develop an intuitive feeling for the matter without `just the right example'. So students may---and perfectly legitimately---ask for a \textit{specific} example conforming to all of their terms\footnote{
Instructors should be well aware that communication always takes place under the \textit{listener's} terms.
}: that it be either electrostatic or gravitational field, and that it be comprehensible and convincing without any calculations.

No problem at all. Consider, one at a time, a configuration of two point charges from \fig~\ref{fig2}. Let the charges be of equal magnitude and let the two configurations differ by their relative sign: either \mbox{$q_1=-q_2$} or \mbox{$q_1=q_2$}. We will call a configuration of opposite signs an `attractive dipole' (\fig~\ref{fig2}, left panel). \textit{Though the configuration of equal signs} (\fig~\ref{fig2}, right panel) \textit{is decidedly not dipolar in nature} (it even features a monopole moment), for clarity of comparison between the two cases we will call it a `repulsive dipole'. We place the two charges on the $z$-axis, at \mbox{$\mathbf{r}_1=+L\,\hat{\mathbf{z}}$} and \mbox{$\mathbf{r}_2=-L\,\hat{\mathbf{z}}$}. Their combined electrostatic field is then:
\begin{equation}
\mathbf{E}_{\pm}(\mathbf{r})=\frac{q_1}{4\pi\epsilon_0}\left(\frac{\mathbf{r}-\mathbf{r}_1}{|\mathbf{r}-\mathbf{r}_1|^3}\pm \frac{\mathbf{r}-\mathbf{r}_2}{|\mathbf{r}-\mathbf{r}_2|^3}\right),
\label{E_dipole}
\end{equation}
with the plus sign denoting a `repulsive dipole' and the minus sign denoting an `attractive dipole'. No calculations with these fields will be necessary, since we count on a general shape of their field lines to be well known.

Let us first examine an `attractive dipole' case from the left panel in \fig~\ref{fig2}. For specificity, let $q_1$ be positive, and $q_2$ be negative. Consider a field at any point along the \textit{symmetry plane} of this setup. For \mbox{$\mathbf{r}_{1,2}=\pm L\,\hat{\mathbf{z}}$} this is a plane defined by \mbox{$z=0$}, indicated by a horizontal dashed line. Either from \eq~(\ref{E_dipole}) or from general symmetry considerations it is abundantly clear that an electric field along this plane is perpendicular to it: \mbox{$\mathbf{E}_-(z=0)\propto -\hat{\mathbf{z}}$}. Now place a positively charged particle between the charges~$q_1$ and~$q_2$, closer to a positive charge~$q_1$, and somewhere off a dipole axis. Specifically, place it at some initial cylindric coordinates \mbox{$\rho_0,z_0$} satisfying \mbox{$\rho_0>0$} and \mbox{$0<z_0<L$}, and let it be accelerated \textit{from rest} by an electrostatic force. Figure~\ref{fig2} shows the field lines due to an electric field~$\mathbf{E}_-$ of an `attractive dipole' \textit{alone}, abstracting it from an electric field of a positive test particle, since it is precisely a field~(\ref{E_dipole}) that affects the motion of a test particle (its trajectory not being affected by its own field). For visual clarity the orientation of field lines from a positive to a negative charge is not indicated; it should be understood that an electric field along each field line is directed from the upper (positive) charge toward the lower (negative) charge. The initial position of a positive test particle is indicated by a black dot off a dipole axis (it is completely irrelevant that it starts from one of the displayed field lines), while a thick oriented line shows its trajectory. The argument for a dissimilarity between the field lines and trajectories is the following. \textit{Every field line is perpendicular to the symmetry plane (everywhere along the same plane) because an electric field is perpendicular to it.} Above the symmetry plane (at \mbox{$z>0$}) the positive test particle is continually pushed \textit{away} from the upper, positive charge: away from the dipole axis and toward the symmetry plane. When it reaches the symmetry plane, it has necessarily attained a \textit{lateral} velocity component (parallel to the plane) due to a continually present force component away from the dipole axis. \textit{Hence, such trajectory can not be perpendicular to a symmetry plane.} This is in striking opposition with any field line piercing the symmetry plane.

A similar argument can be made with the well known field lines of a `repulsive dipole' configuration. For specificity, let both charges~$q_1$ and~$q_2$ be positive. Right panel from \fig~\ref{fig2} shows the undirected field lines due to a `repulsive dipole' field~$\mathbf{E}_+$ alone. (The cross at the origin marks that the field vanishes there and the field lines terminate there: the vertical field lines approach the termination point, the horizontal ones depart from it.) Let the positive charged particle again be placed between~$q_1$ and~$q_2$---closer to the upper positive charge $q_1$ and \textit{very close to the dipole axis}---and let it accelerate from rest due to a field~$\mathbf{E}_+$. First imagine what would happen if a test charge was placed precisely on the dipole axis. It would oscillate between~$q_1$ and~$q_2$ (reaching a minimal distance from~$q_2$ equal to the initial distance from~$q_1$), whereby it would certainly be crossing the symmetry plane \mbox{$z=0$}. Now gradually shift the initial test charge position off the dipole axis. Due to the continuity of trajectory variations with varying initial position (i.e. a continuity of a family of trajectories), for slight offsets from the dipole axis (infinitesimal at first) a test charge \textit{will still be crossing a symmetry plane, having a velocity component perpendicular to it}. On the other hand, from the symmetry of a `repulsive dipole' setup it is evident that $\mathbf{E}_+$ has only a lateral component along the symmetry plane: \mbox{$\mathbf{E}_+(z=0)\propto \hat{\boldsymbol{\rho}}$}. \textit{Thus, unlike the particle trajectory, no field line can cross the symmetry plane}.

We consider it important for instructors to be able to lead a previous discussion without much calculations. For this reason we have presented counterexamples where the field lines themselves are obvious (the circular ones from \fig~\ref{fig1}) or so well known and widespread in literature (`dipole' field lines from \fig~\ref{fig2}) that one can draw them by hand, even without precise mathematics. Instructors who can afford to go into further details may want to take note of the general field line equation (not as a particular-field-line parametrization, but rather a general differential equation yielding the particular field line solutions), presented in Appendix~\ref{appendix_A}.

\vspace*{-2mm}

\section{Field lines density as a measure of a field magnitude}
\label{density}

There is a claim going around that a density of field lines represents a magnitude of a vector field. This misconception is remarkably entrenched in literature, \textit{in this simplistic form}. In fact, it does not even seem to be so widespread among students, at least not until this fallacious idea has been sufficiently forced upon them. Before commenting further on the status and nature of this misconception, let us immediately give a trivial counterexample to the claim.

\begin{figure}[t!]
\centering
\includegraphics[height=\figheight\textwidth,keepaspectratio]{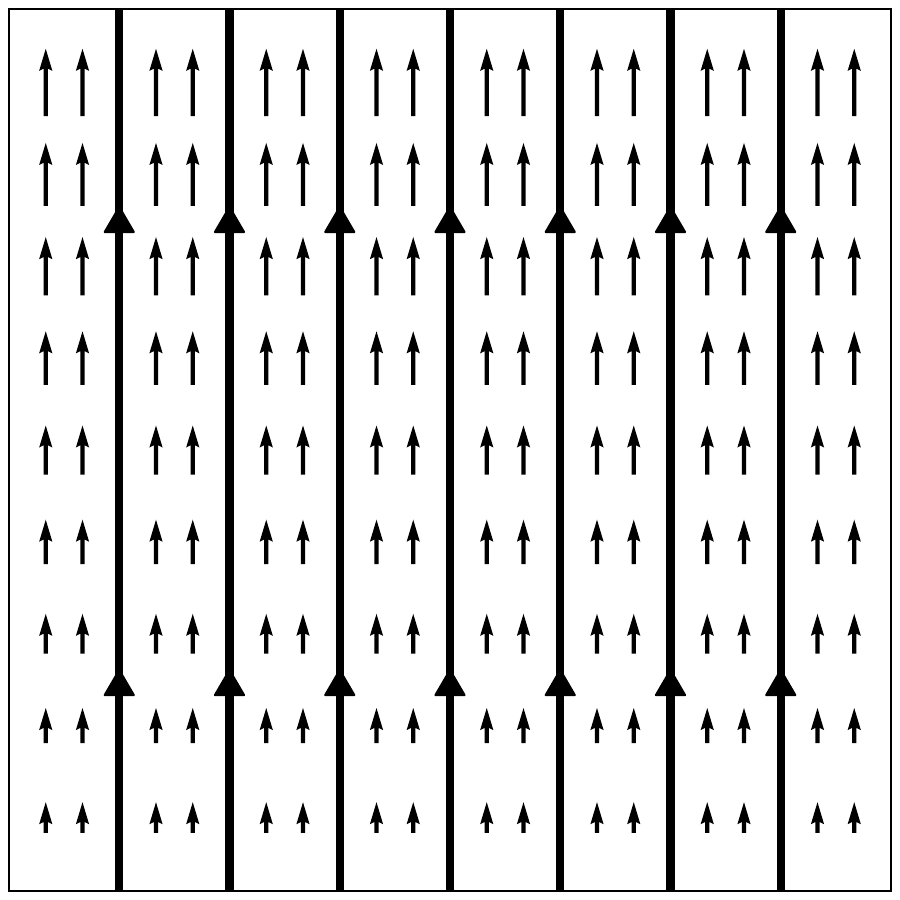}
\caption{Counterexample to a notion that a density of field lines measures a field magnitude. Several field lines (solid directed lines) are shown together with a generating vector field of uniform direction and linearly increasing magnitude: \mbox{$\mathbf{F}(\mathbf{r})=f_0 x\,\hat{\mathbf{x}}$} ($x$-coordinate corresponding to a vertical axis).}
\label{fig3}
\end{figure}

Consider a homogeneous vector field, everywhere equal in direction and magnitude. For example: \mbox{$\mathbf{F}(\mathbf{r})=F_0\,\hat{\mathbf{n}}$}, with~$\hat{\mathbf{n}}$ as some fixed unit direction. How do its field lines look and how are they distributed? Now consider a field of\linebreak the \textit{same direction} but increasing magnitude, like the one presented in \fig~\ref{fig3}. How does this affect the appearance and the density of field lines? The answer is painfully obvious: \textit{it does not; not in the slightest}. In both cases all field lines are exactly the same. They consist of the mutually parallel straight lines. Their density is entirely unaffected relative to a homogeneous field, even though a magnitude of a vector field has been altered. In fact, its magnitude could vary \textit{arbitrarily} from point to point:
\begin{equation}
\mathbf{F}(\mathbf{r})=F(\mathbf{r})\hat{\mathbf{n}},
\label{uniform}
\end{equation}
and the field lines would still remain unaffected\footnote{
There is a limitation to this arbitrariness: a pointwise dependence $F(\mathbf{r})$ should be strictly positive: \mbox{$F(\mathbf{r})>0$}. Otherwise, a given field line would have to stop (would suddenly be cut off) where a field magnitude vanishes or abruptly changes sign.
}.

With this counterexample in mind we cite several widely used books perpetuating the initial claim~\cite{book_serway,book_young,book_jones,book_halliday,book_knight}. This list is far from exhaustive and is limited only by our own exhaustion in composing it.  In fact, some sources go so far as to state that the density of the electric field lines is \textit{by definition} directly proportional to a magnitude of an electric field~\cite{students_0,book_kenworthy}.

Even the authors of many dedicated studies into students' understanding of the field line concept and their struggles with it perpetuate this misconception~\cite{lines_force,trajectory_0,trajectory_3,students_0,students_1,students_2,students_3}. Some only skim over the claim, presenting it as one of the `basic facts' about the field lines. Others expect the students to demonstrate their knowledge and understanding by regurgitating and actively using the `fact'. This is then translated into a flawed assessment of some students' answers, wherein otherwise correct answers are interpreted as incorrect and vice versa\footnote{
Refs.~\cite{students_2,students_3}, for example, use a field from \fig~\ref{fig3}---\textit{a very counterexample to a relation between the field lines and field magnitude}---and expect the students to demonstrate a `mastery' of the misconception by representing a field magnitude by a varying field line density. The students' failure to do so is then found to be the \textit{main difficulty} the students have with the field lines~\cite{students_3}.
}. Thus, a failure at falsity is indeed treated as a failure.

In light of most of these claims focusing specifically on an electric field, it is natural to ask if the field from our counterexample (\fig~\ref{fig3}) may indeed be an electric field. Not only is the answer affirmative, but the electric field in question is the one regularly used in pedagogical demonstrations of the applicability of Gauss's law---it is a field of an infinite charged slab, inside the slab itself. For a slab of uniform charge density~$\varrho_0$, extending between $x=\pm L$, an electric field~$\mathbf{E}(\mathbf{r})$ is:
\begin{equation}
\mathbf{E}(\mathbf{r})=\left\{\begin{array}{lcc}
(x\varrho_0 /\epsilon_0) \hat{\mathbf{x}} &\mathrm{for}& |x|\le L,\\
(L\varrho_0 /\epsilon_0) \hat{\mathbf{x}} &\mathrm{for}& |x|> L,\\
\end{array}\right. 
\label{E_slab}
\end{equation}
with $\epsilon_0$ as the permittivity of vacuum. That a field from \fig~\ref{fig3}---found within a slab---does not extend over an entire space is not a valid  complaint. In order to disprove the initial claim about the field lines density, one only needs to find a portion of space where the claim does not hold. This example fully achieves this goal. Those reluctant to be convinced may also protest that this particular field is not perfectly achievable in practice, due to the requirement for an infinite slab. This argument is also null and void, as an electric field~(\ref{E_slab}) \textit{does} satisfy Maxwell's equations, i.e. the basic laws of electrodynamics, and therefore \textit{is} by definition a valid electric field. After all, one does not need a precise field from \fig~\ref{fig3} or even a more general one from \eq~(\ref{uniform}). Even some approximation of these fields is sufficient to be a convincing counterexample; as soon as a field magnitude sufficiently deviates from uniform while a field direction stays barely affected, the initial erroneous claim is disproved. In that, electric fields approximating a relevant portion of the field~(\ref{E_slab}) can certainly be realized in practice, sufficiently deep inside a finite slab of electrically isolating material.

\begin{figure}[t!]
\centering
\includegraphics[height=\figheight\textwidth,keepaspectratio]{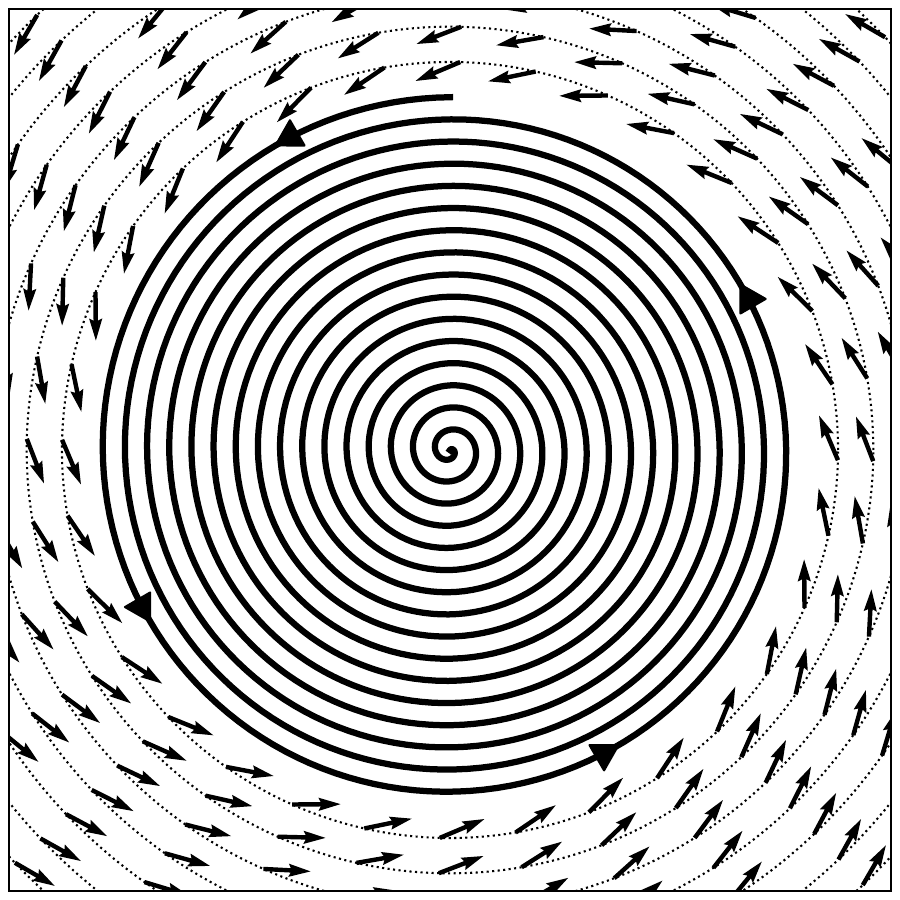}
\caption{Example illustrating that the field line density behavior does not correlate with the field magnitude behavior, when altering the field parameters.
A spiraling curve is a single field line of an electric field~(\ref{E_spiral}). For illustration purposes it is artificially terminated and a generating field itself is illustrated by a vector plot outside a displayed field line~range.}
\label{fig4}
\end{figure}

There is a beautiful example by Herrmann et al.~\cite{electric_magnetic}, further illustrating the dangers in assessing the field magnitude behavior---in particular its dependence on the field parameters---from the field line diagrams. They consider an electric field having both a radial and azimuthal component, which we slightly alter for our demonstration\footnote{
Both components can be produced by a long (ideally infinite) wire, carrying a static and uniform line charge distribution and a current varying linearly in time. For example, a conductive wire with a surplus of electrons that provide both a total charge and a required current. In that, a radial field component is produced by a uniform charge distribution, while the azimuthal component is produced by magnetic induction (the Faraday's law), from a surrounding magnetic field that linearly varies in time.
}:
\begin{equation}
\mathbf{E}(\mathbf{r})=\frac{\varepsilon}{\rho}\,\hat{\boldsymbol{\rho}}+\mathcal{E}\,\hat{\boldsymbol{\varphi}},
\label{E_spiral}
\end{equation}
and express it in cylindrical coordinates (\mbox{$E_\rho=\varepsilon/\rho$} and \mbox{$E_\varphi=\mathcal{E}$}). Even without any calculations it is evident that the field lines will spiral around the $z$-axis, as a source of a radial component. Figure~\ref{fig4} shows an example for \mbox{$\varepsilon<0$}, yielding a `collapsing' spiral. Now consider what happens when we keep the azimuthal parameter~$\mathcal{E}$ constant, while \textit{reducing} the magnitude of the radial parameter~$\varepsilon$. As \mbox{$\varepsilon\to 0$}, a field magnitude decreases, while a spiral becomes denser and denser. After each full circle it passes closer and closer to a starting point, so that by varying~$\varepsilon$ a field line density\footnote{
Though an absolute field line density around any point is infinite, one can meaningfully talk about a \textit{relative} density between different points. See the Supplementary note for further details.} \textit{increases} with a decreasing field magnitude! So much so that as $\varepsilon$ approaches zero, a field line density diverges, becoming \textit{infinite}! Thus, a field line density behaves in a \textit{completely opposite} manner to a field magnitude, failing to even approximate its dependence on the field parameters by any stretch of the imagination. It should be well noted that the relation examined here is not the same relation as before---that between the field line density and a field magnitude \textit{throughout a given field}---but rather a relation between field lines for different fields. In particular, how the field line density correlates with a field magnitude, not throughout space but in regard to the field parameters.

\pagebreak

\vspace*{-5mm}

Yet, this example \textit{also} serves to disprove the initially addressed relation. First observe that the distance between any two successive spiral coils from \fig~\ref{fig4} seems to be constant in the radial direction. And indeed it is---its field line equation may be found in later table~\ref{tab1}: the one with a \textit{linear} dependence for~$\rho(\varphi)$. Now notice that between any two successive coils of any single field line \textit{all other field lines are contained}! This means the density of field lines---of all of them and not only of a single one---is uniform in the radial direction. On the other hand, the magnitude of a field~(\ref{E_spiral}) evidently decreases in the same direction. Hence, this is another counterexample to a general claim about a direct relation between a field line density and a field magnitude---an electric field at that!\linebreak

\vspace*{-3.5mm}

While clearly recognizing that a relation between a field magnitude and a field line density is generally untenable, some authors argue that it could be preserved by introducing artificial cuts, i.e. by arbitrarily interrupting the field lines and starting the new ones as necessary~\cite{magnetic_1,redundant}. This is, of course, a logically unfounded procedure, equivalent to maintaining a fixed idea that must be preserved at any price, even at the cost of introducing a nonsense or a contradiction. And the proposed procedure is, indeed, a semantic nonsense. It consists of a logical fallacy \textit{equivocation}: pronouncing and thus treating the field line \textit{segments} as the field lines themselves. By definition, a field line is a \textit{whole} solution to a field line equation~(\ref{master}); it is either an uninterrupted curve, or one terminated only by a vanishing or undefined field. A portion of a circle is no longer a circle but a circular arc. A square cut in half is no longer a square but a rectangle, a trapezoid or a triangle. A sphere appropriately chopped off so that a cube remains is no longer a sphere but a cube. So is an arbitrary field line cutaway no longer a field line but a \textit{field line segment}. In light of earlier counterexamples, a general claim could be made only to the effect that an \textit{arbitrary selection of arbitrary field line segments} might possibly hint at a field magnitude. But hardly anyone seems to be making \textit{that} claim.

\pagebreak

Not to be unfair, we find the original proposal by Slepian~\cite{redundant} and McDonald~\cite{magnetic_1} for terminating the field lines to be a mere `soft' suggestion intended to provide at least approximate means of assessing a field magnitude. This is clearly reflected by the authors themselves recognizing that the proposed field line interruptions are artificial and arbitrary. In fact, the cited authors are among the earliest champions attempting to dispel the widespread field line misconceptions. However, such well intended ideas are all too easily taken too far by the uninitiated.

In the end, we note that most pedagogical sources discussing the field lines never give an actual field line equation [\eq~(\ref{master}) from Appendix~\ref{appendix_A}], either in a form~(\ref{master}) or in any other equivalent form (see~\cite{our2}). We are not the first to notice this~\cite{electric_1}. It is no wonder then that field line misconceptions abound, since the field lines are rarely defined (except by vague verbal descriptions) as precise mathematical objects. Only a few limited field line examples are ever shown. General field line properties are then inductively extrapolated from these examples that `everybody knows', while no other example can be explored due to a lack of their clear mathematical formulation. Unfortunately, even those who do give some form of the general field line equation often claim a connection between a field line density and a field magnitude as a `known fact'. This is because the forms of a field line equation other than \eq~(\ref{master}), while mathematically equivalent, do not reveal in an obvious way what is painfully visible from \eq~(\ref{master}). Since a \textit{unit} field $\hat{\mathbf{F}}(\mathbf{r})$ appears in it, the general field line equation is entirely independent of and insensitive to a field magnitude! Thus, neither the field lines themselves nor any single aspect of theirs can possibly be affected by a field magnitude. A given field can always be arbitrarily rescaled not by a single number, but by an entire (strictly positive but otherwise arbitrary\footnote{
That a rescaling function $f(\mathbf{r})$ should be nonzero, at least where $\mathbf{F}(\mathbf{r})\neq\mathbf{0}$, is obvious. Otherwise, the field lines of a rescaled field $f(\mathbf{r})\mathbf{F}(\mathbf{r})$ would end at points where \mbox{$f(\mathbf{r})=0$}, in opposition with the field lines of the original field $\mathbf{F}(\mathbf{r})$. The same would occur if $f(\mathbf{r})$ discontinuously switched sign.
}) scalar function: \mbox{$\mathbf{F}(\mathbf{r})\to f(\mathbf{r})\mathbf{F}(\mathbf{r})$}, and its field lines remain perfectly unaffected. \textit{Only a field direction counts!} A field line density may turn out to be related to a field magnitude only by serendipitous chance in selecting a particular field example.

Just as in Section~\ref{trajectories}, we strive to isolate the technical details from the main discussion, since we wish for instructors to be convincing without having to resort to a `heavy artillery'. However, as the avoidance of the general field line equation [\eq~(\ref{master}) from Appendix~\ref{appendix_A}] is part of a problem, a demonstration of its application is in order, which is provided in Appendix~\ref{appendix_B}. 

\vspace*{-5mm}

\subsection*{Divergence-free fields}

\vspace*{-3mm}

It is sometimes argued that a field line density represents a field magnitude where a field divergence vanishes (\mbox{$\boldsymbol{\nabla}\cdot\mathbf{F}=0$}), as a consequence of Gauss's law~\cite{redundant}. However, this claim needs to be assisted by several technical caveats, that are rarely given. One caveat---in our experience generally known, but rarely stated---is that since the familiar Gauss's law holds in a three-dimensional space, only a three-dimensional configuration of field lines could stand a chance of accurately representing a field magnitude. Thus, planar field line representations (as two-dimensional drawings on a screen or paper) immediately fail at \textit{even approximately accurate} representation of a field magnitude. It is well known that an electric field of a point charge would have to vary as \mbox{$1/r$}, instead of \mbox{$1/r^2$}, in order for Gauss's law to hold in two dimensions. Furthermore, planar representations of a familiar $1/r^2$ field generally suffer from dangerously misleading visual artifacts such as a false monopole moment~\cite{electric_2,electric_3}, equatorial clumping and boundary clumping~\cite{electric_3} (we analyze some of these examples in the Supplementary note). These effects simply make it impossible to draw accurate conclusions about a field magnitude from the two-dimensional field line diagrams.

The other caveat seems to be lost to history, as we are aware of only one single reference---by Wolf et al.~\cite{electric_3}---even mentioning it, let alone properly exposing it. From Gauss's law they derive a \textit{proper} relation between the field line density and a field magnitude. Instead of repeating their derivation here, we refer the reader to their original work~\cite{electric_3}, as it is one of the most important and insightful references on the subject. Let \mbox{$\varrho(\mathbf{r},\hat{\mathbf{n}})$} be the local density of field lines piercing an infinitesimal, oriented surface element \mbox{$\D\mathbf{A}=\D A \,\hat{\mathbf{n}}$}, its orientation defined by a normal vector~$\hat{\mathbf{n}}$. Wolf et al. show that---for a field satisfying Gauss's law---the field line density is \textit{not} proportional to a field magnitude, but rather to a field component perpendicular to an observed surface element:
\begin{equation}
\varrho(\mathbf{r},\hat{\mathbf{n}})\propto \mathbf{F}(\mathbf{r})\cdot \hat{\mathbf{n}},
\label{rho_relation}
\end{equation}
i.e. to a `field magnitude in a direction of ~$\hat{\mathbf{n}}$'. This is by no means a trivial claim, neither by its familiarity (or the lack thereof) nor its content. Regarding its familiarity, even a requirement for a vanishing divergence itself (\mbox{$\boldsymbol{\nabla}\cdot\mathbf{F}=0$}) is hardly ever mentioned. Even Slepian, who clearly brings it up---and who seems to have started in the early 1950s a 'movement' against the field line misconceptions~\cite{redundant}---says: `[...] if the divergence of the vector field is zero there, we may prolong the same continuous lines of force away from the region, and the assemblage of these prolonged lines of force will continue by its density to represent the \textit{magnitude of the vector field}' (italics our own).

Regarding a nontriviality of the content of \eq~(\ref{rho_relation}), consider an earlier spiraling field line from \fig~\ref{fig4}, associated with a field~(\ref{E_spiral}). This field is indeed divergence-free, \textit{everywhere} except at the $z$-axis (where it is described by a Dirac delta function: \mbox{$\boldsymbol{\nabla}\cdot\mathbf{E}=\varepsilon \delta(\rho)/\rho$}). In the context of \fig~\ref{fig4} we have already seen that a field line density, which is  \textit{uniform in a radial direction}, does not correlate to a field magnitude, which decreases in the same direction. However, a field component perpendicular to the inspected radial direction is an \textit{azimuthal} field component \mbox{$E_\varphi=\mathcal{E}$}, which is indeed constant, in full agreement with \eq~(\ref{rho_relation}). We demonstrate some further examples of the validity of \eq~(\ref{rho_relation}) in the Supplementary note.

If one were to obtain any sense for a total-field magnitude from an assemblage of field lines, one would---in accordance with \eq~(\ref{rho_relation})---have to consider a field line density in \textit{three independent directions}, in order to assess three independent field components. In conventional field line diagrams, consisting of planar field line representations, one of these components is by definition inaccessible! In light of all these insights, Wolf et al. in their forcefully but aptly named paper `Electric field line diagrams don't work' conclude~\cite{electric_3}:\\

\begin{addmargin}[1em]{1em}
`In fact, in almost every [conventional field line diagram], \textit{field strength has no consistent relationship to the observed field line density}.' \\
\end{addmargin}

Having touched upon Gauss's law, we return shortly to the issue of arbitrary and artificial termination of the field lines. If one were to plead this proposal for the divergence-free fields---on a basis that Gauss's law makes at least \textit{some} aspects of a field line density related to \textit{some} aspects of a field magnitude---the same law would immediately render the proposal logically inconsistent. Gauss's law associates a number of field lines to a field flux. Surely, this means that the field line terminations act as the \textit{sinks} or \textit{sources} of a flux. Hence, within this mental framework, introducing the field line interruptions is equivalent to introducing nonexistent charges, electric or otherwise!

In conclusion, a field line density \textit{is} an objective property of the field lines, as they are well defined mathematical objects. However, a field line density is demonstrably and decidedly \textit{not} related to a vector field magnitude, except incidentally, by spurious correlation. Therefore, \textit{this relation can not be forced either by convention or definition}. In general, not even our own particular selection of \textit{displayed} field lines or the field line segments reflects a vector field magnitude, neither accurately nor consistently. Instead, a density of displayed field lines reflects only \textit{our own selection of their starting points}.

There \textit{is} something, a density of which \textit{is} directly related to a vector field density. These are the \textit{equipotentials} of a scalar field $f(\mathbf{r})$ derived from a vector field $\mathbf{F}(\mathbf{r})$, such that \mbox{$\mathbf{F}(\mathbf{r})=(\pm)\boldsymbol{\nabla}f(\mathbf{r})$}. It follows from \textit{this} definition that a magnitude of a vector field between two close scalar equipotentials \textit{is} determined by their closeness. In rough terms: \mbox{$F\approx \Delta f/\Delta \ell$}, with $\Delta \ell$ as a local distance between two equipotentials whose scalar potentials differ by~$\Delta f$. However, a scalar potential can be derived and its equipotentials can be defined \textit{only} for the conservative, i.e. rotationless vector fields (\mbox{$\boldsymbol{\nabla}\times\mathbf{F}=\mathbf{0}$}). Another complication is that---unlike the field lines---equipotentials may form subspaces (manifolds) of any dimensionality: they may be points, curves, surfaces, solids or hyper-solids of the same dimensionality as an embedding space, or may even \textit{be} an entire embedding space (consider the  equipotentials of a vanishing vector field \mbox{$\mathbf{F}=\mathbf{0}$}). Thus, any discussion about a density of equipotentials is necessarily less `clear cut' than wishful thinking about a density of field lines. However, this does not make a demonstrably fallacious wishful thinking a valid surrogate. A moral could hardly be stated more clearly than in~\cite{electric_magnetic}:\\ 

\begin{addmargin}[1em]{1em}
`Take advantage of the virtues of conventional field line
diagrams but don't pretend they represent something that they do not. In other words: Don't claim that the line density measures the field strength.'
\end{addmargin}

\section{Closed field lines of a divergence-free field}
\label{closed_lines}

The final misconception we address is commonly encountered as a claim that the field lines of a magnetic field always form closed loops. Its source is basically the same as a source of misconception about a field line density: it consists of an incomplete induction based on a minimal set of simplistic examples. These examples usually involve a magnetic field of: (1)~an infinitely long straight wire carrying a steady current, (2)~a perfectly circular wire carrying a steady current, (3)~a geometrically regular permanent magnet or a single magnetic point dipole\footnote{
Even a circular wire setup and a regular dipole magnet provide a counterexample to the closed field lines claim: a single field line lying along the axis of the axial symmetry. It is a straight line, therefore \textit{open at both ends}. It can not be considered to ``loop around'' at an infinite distance from the setup due to the uniqueness problem: in which azimuthal direction should such field line ``turn around'' at infinity, when all directions are equivalent?
}. Among other logical errors working in tandem with these examples, the most prominent one is a supposed connection between the magnetic field lines and a magnetic flux, wherein a vanishing field divergence (\mbox{$\boldsymbol{\nabla}\cdot \mathbf{B}=0$}) does indeed have implications upon the flux itself.

To this day ample literature has been produced showing beyond any trace of a doubt that magnetic field lines do not, in general, form closed loops~\cite{redundant,magnetic_1,magnetic_2,magnetic_3,magnetic_4,ajp_new}. In fact, they \textit{almost never do}! As soon as a particular magnetic field source deviates from an excessive simplicity of a few given examples, one invariably finds a claim about the closed magnetic field lines to be false. A simple, familiar and oft quoted example is a magnetic field produced by two simultaneous currents: one flowing along a conductive ring, the other along a long wire placed along the ring axis~\cite{magnetic_1,magnetic_3,magnetic_4,ajp_new}. A specially beautiful treatment of this simple setup can be found in~\cite{magnetic_4}.

Why then is this misconception still `in circulation'? Because the amount of literature still perpetuating it far surpasses all the work aimed at dispelling it. The very first sentence from~\cite{magnetic_1} states: `Textbooks often hold erroneously that the condition \mbox{$\boldsymbol{\nabla}\cdot \mathbf{H}=0$} implies that lines and tubes of force are closed.' An extensive overview of such textbook examples may be found in~\cite{magnetic_3,magnetic_4}. In certain cases the `fact' of the closed magnetic field lines is implicitly suggested by erroneously presented examples~\cite{magnetic_3}. At other times the `fact' is explicitly stated and specially emphasized (e.g. Chapter 27.3 from~\cite{book_young}).

A vanishing divergence of a magnetic field means that no spatial point can act as a net source or sink of a magnetic flux. By relating the magnetic field lines with a magnetic flux, it follows in a few `logical' bounds (a self-flattering term for flexible mental acrobatics) that field lines can have no ends and therefore must form closed loops. The `original sin' in this thinking is, of course, attributing a level of physical reality to the field lines by relating them in any way to a magnetic flux. We have already seen that this connection is untenable, by being able to construct a desired field line density entirely independent of a field magnitude (which, of course, affects the flux). Hence, any conclusion drawn from such erroneous premise is invalid. But is it wrong? Through a disconnect with a physical reality any conclusion based on a meaningful physical quantity is immediately rendered \textit{not necessarily} true. An option then remains that a claim about the magnetic field lines as closed loops might be \textit{incidentally} true, for some other reason.

In order to quickly demonstrate that a vanishing field divergence has nothing to do with the closed field lines, one only needs to consider a homogeneous field: \mbox{$\mathbf{F}(\mathbf{r})=F_0\,\hat{\mathbf{n}}$} (both~$F_0$ and~$\hat{\mathbf{n}}$ being constant). All field lines are straight lines, thus open curves. At this point one can complain that this field is rotationless, unlike the magnetic fields. In this case just consider a `twisted' field \mbox{$\mathbf{F}(\mathbf{r})=F_\varphi\hat{\boldsymbol{\varphi}}+F_z\hat{\mathbf{z}}$}, both~$F_\varphi$ and~$F_z$ being constant, so that \mbox{$\boldsymbol{\nabla}\cdot\mathbf{F}=0$} and \mbox{$\boldsymbol{\nabla}\times\mathbf{F}=(F_\varphi/\rho)\hat{\mathbf{z}}$}. Its field lines are \textit{obviously} spirals; once again open curves\footnote{
One might still complain whether these could be magnetic fields. Certainly, though only within  a limited potion of space, which is entirely sufficient to disprove a misconception of \textit{all} magnetic field lines as closed loops. In fact, a homogeneous magnetic field inside an infinitely long, densely wound solenoid is a textbook example that everyone uses as some point. Yet, rarely anyone recognizes it as a trivial counterexample to the closed field lines claim. A proposed rotational field should also be contained within a limited portion of space in order to pronounce it a magnetic field: \mbox{$\mathbf{B}(\mathbf{r})=B_\varphi\hat{\boldsymbol{\varphi}}+B_z\hat{\mathbf{z}}$}. A current density~$\mathbf{J}$ producing it inside an infinitely long cylinder of radius~$\rho_0$ may be modeled using a Dirac delta function: \mbox{$\mu_0\mathbf{J}=B_z\delta(\rho-\rho_0)\hat{\boldsymbol{\varphi}}+(B_\varphi/\rho)\hat{\mathbf{z}}$}. 
}.

Most of the counterexamples present in literature~\cite{redundant,magnetic_1,magnetic_2,magnetic_3,magnetic_4}---together with the two proposed here---aim to disprove a claim about the closed field lines by demonstrating that they may be open curves \textit{without ends}, extending indefinitely. But we can do much better than that! Being open and without ends is indeed \textit{sufficient} for the magnetic field lines not to be closed. But is it \textit{necessary}? The answer is a resounding \textit{no}~\cite{ajp_new} and a reason for it is as ingenious as it is simple\footnote{Though simple, the answer is by no means `cheap', as hardly anyone seems to be aware of it. In fact, the idea seems to have originated as a byproduct of a rather complex analysis in~\cite{magnetic_1}.
}. Can a magnetic field vanish at any point (\mbox{$\mathbf{B}=\mathbf{0}$})? Of course! So what happens with the field lines at this point? They terminate! Depending on a field direction around this point, the field lines either stop or originate at it. In any case, they certainly do not close a loop. 

\pagebreak

\begin{figure}[b!]
\centering
\vspace*{-2mm}
\includegraphics[width=0.8\linewidth,keepaspectratio]{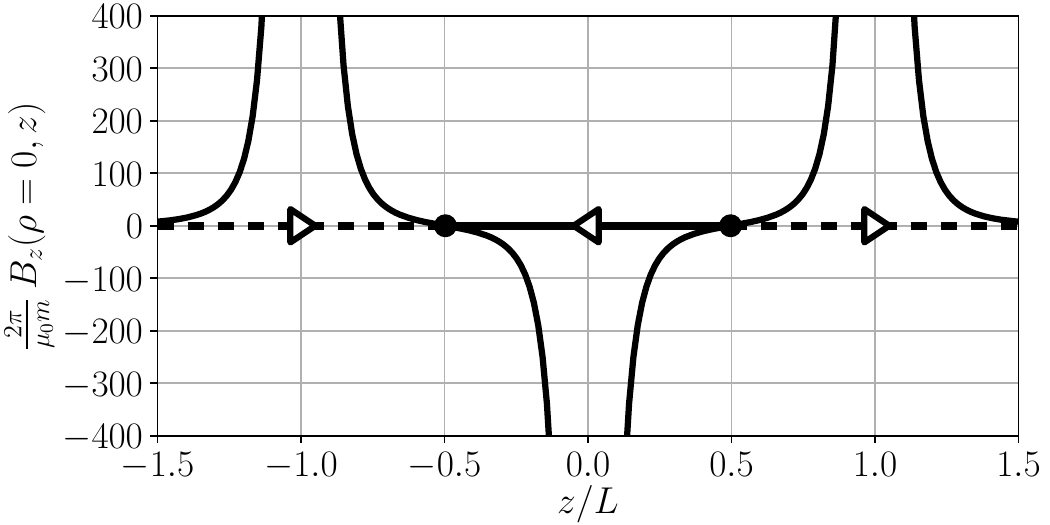}
\caption{$B_z$-component of a magnetic field from \eq~(\ref{field_bz}). Two null-points (black dots) are located at \mbox{$z/L\approx\pm0.497$}. Field lines lying fully along the $z$-axis are also indicated. Two dashed field lines terminate at one end and extend indefinitely at the other. A solid field line terminates at both ends.}
\label{fig6}
\end{figure}

\begin{figure*}[t!]
\centering
\includegraphics[height=\figheight\textwidth,keepaspectratio]{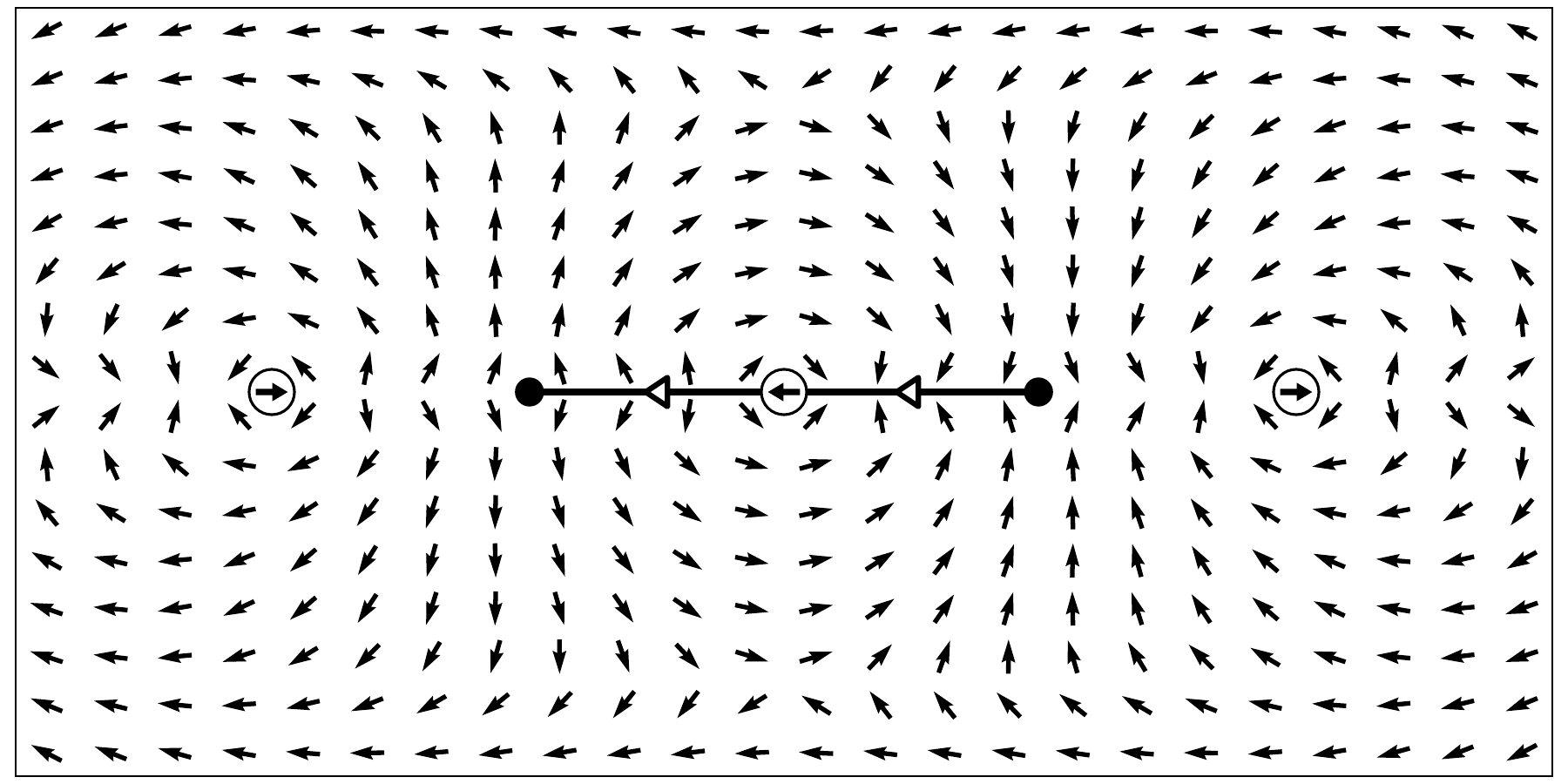} 
\caption{Counterexample to a claim that magnetic field lines always form closed loops. Three point dipoles (circled arrows) are carefully arranged so as to produce a magnetic field yielding a field line terminating at both ends.
}
\label{fig7}
\vspace*{5mm}
\end{figure*}

We will now provide a trivial and convincing example showing that the magnetic field lines may terminate \textit{at both ends}. A success in doing so should put to rest any and all speculations about them always being closed, or even `just' not having ends. For this we need a setup with at least two separate points of a vanishing magnetic field. In addition, we need it to be perfectly clear and obvious---without solving any equations, differential or otherwise---that there exists a field line that starts at one of these points and ends precisely at the other one. We can achieve this if \textit{everywhere along the line connecting the two null-field points}, a magnetic field is collinear to the line connecting these points. The field line starting at one null-field point would then be guided directly toward the other, where it would terminate. The null-field points themselves can be realized by a destructive superposition of magnetic fields from separate sources; the simplest realization being the field contributions of opposing directions. These requirements strongly hint at a configuration of three magnetic dipoles, \textit{placed and directed} along a single line. In that, the middle dipole should be oriented in the opposite direction from the two outer ones (which should be oriented in the same direction). This layout provides a destructive field superposition we are aiming for. We will consider a field of three point dipoles\footnote{
The relevant part of the construction would also be achieved by using the finite-size dipoles, i.e. geometrically regular magnets of finite magnetization density. The only advantage to using the point dipoles is a simple expression for the magnetic field from \eq~(\ref{B_point_dipole}), which would be more complicated for finite-size dipoles.
}. A field contribution from a single ($i$-th) magnetic point dipole of a dipole moment $\mathbf{m}_i$, located at~$\mathbf{r}_i$, is:
\begin{equation}
\mathbf{B}_i(\mathbf{r})=\frac{\mu_0}{4\pi}\frac{3(\mathbf{m}_i\cdot\hat{\mathbf{R}}_i)\hat{\mathbf{R}}_i-\mathbf{m}_i}{R_i^3},
\label{B_point_dipole}
\end{equation}
with \mbox{$\mathbf{R}_i=\mathbf{r}-\mathbf{r}_i$}, so that \mbox{$R_i=|\mathbf{R}_i|$} and \mbox{$\hat{\mathbf{R}}_i=\mathbf{R}_i/R_i$}. For simplicity we consider two outer dipoles at the same distance from the middle one, and place them all along the $z$-axis. We index them according to their \mbox{$z$-coordinate}, so that \mbox{$\mathbf{r}_\pm=\pm L\,\hat{\mathbf{z}}$} and \mbox{$\mathbf{r}_0=\mathbf{0}$}, and select their dipole moments as \mbox{$\mathbf{m}_\pm=m\,\hat{\mathbf{z}}$} and \mbox{$\mathbf{m}_0=-m\,\hat{\mathbf{z}}$}. The total magnetic field of such configuration is simply:
\begin{equation}
\mathbf{B}(\mathbf{r})=\mathbf{B}_+(\mathbf{r})+\mathbf{B}_0(\mathbf{r})+\mathbf{B}_-(\mathbf{r}).
\label{field_total}
\end{equation}

\noindent As we are mainly interested in the field along the $z$-axis, we write it out in full (a dependence shown in \fig~\ref{fig6}):
\begin{equation}
\mathbf{B}(\rho=0,z)=\frac{\mu_0 m}{2\pi}\left(\frac{1}{|z-L|^3}-\frac{1}{|z|^3}+\frac{1}{|z+L|^3}\right)\hat{\mathbf{z}}.
\label{field_bz}
\end{equation}
Even without considering this expression, it is clear from the symmetry of the setup that there are indeed two null-field points along the $z$-axis, approximately at \mbox{$\pm(L/2)\hat{\mathbf{z}}$}. By numerical search one finds more accurate (still approximate) roots of \eq~(\ref{field_bz}) to be \mbox{$ \pm0.496895 L$}.

Figure~\ref{fig7} shows a magnetic field~(\ref{field_total}) around the three point dipoles. Black dots are the points of field line terminations, corresponding to the \textit{relevant} null-points of a magnetic field (there is an entire continuous curve of a vanishing magnetic field, surrounding the middle dipole). A displayed field line terminates \textit{at both ends}, disproving in a most decisive manner a claim about the magnetic field lines always forming closed loops\footnote{
A fact that the magnetic field lines may terminate does not suggest, even remotely, that a magnetic field has non-vanishing divergence or that magnetic monopoles exist. A nice and detailed discussion about this may be found in~\cite{magnetic_1}. A side note: in our setup all field lines \textit{in the immediate vicinity} of the relevant null-field points (with the exception of three terminating lines lying on the $z$-axis) locally resemble hyperbolas, as the local magnetic field corresponds to a so-called  \textit{deformation} field from~\cite{magnetic_1}.
}. There are two additional field lines lying on the $z$-axis, as indicated in \fig~\ref{fig6}. Each one terminates at one of the null-points (at one end, while extending indefinitely at the other), thus also providing usable counterexamples. For visual clarity only the central field line is shown in \fig~\ref{fig7}.

In summary, the closed field lines misconception is an extreme case of a `mistaken identity': instead of \mbox{$\boldsymbol{\nabla}\cdot \mathbf{B}=0$} implying that the magnetic field lines should be closed, the possibility of \mbox{$\mathbf{B}=\mathbf{0}$} implies that they should \textit{not} be!

\pagebreak

\section{Conclusions}

We have addressed the most notable field line misconceptions afflicting the students of physical sciences. The historical notion of the field lines as of `lines of force' reaches all the way back to Faraday and his initial conceptions. In time his ideas were refined and today we have a much better understanding of these well defined mathematical objects. Some of these realizations have long since found their way into the pedagogical literature. However, some misconceptions still permeate a vast portion of the contemporary educational sources. A collective mentality still seems to be beset with the ideas of a field line density measuring a magnitude of a vector field, and of magnetic field lines always forming closed loops.

First and foremost, the field lines are hardly anything more than arbitrarily chosen integral curves. There is no physical meaning to their density, continuity or closedness. There is certainly no general connection with a magnitude of a vector field. This is clear both from the general field line equation [Eq.~(\ref{master}) from Appendix~\ref{appendix_A}] which is \textit{completely independent} of a vector field magnitude, and from multiple counterexamples that can be constructed even for the very specific fields (electric, magnetic, gravitational) that are widely considered to have the field lines of `special' properties. All objective field line properties are determined \textit{only} by a direction of a vector field.

In \textit{drawing} the field lines their density is ultimately determined by our own choice of \textit{starting points} for tracing each particular one. In fact, drawing them in a particular manner \textit{can} have its advantages, depending on a message one is trying to convey. For example, drawing the `Lorentz contracted' field lines of `relativistically affected' density is useful for illustrating a relativistic transformation of an electric field, especially that of a point charge in uniform motion~\cite{our2,our1}. However, the purpose behind such procedure needs to be well understood so that no hidden and unintended meaning is attributed to what is otherwise an arbitrary choice.

\pagebreak

Throughout the main body of this paper, the Appendices and the associated Supplementary note we have demonstrated multiple counterexamples and levels of argumentation against each addressed type of misconception. As a concise overview of the salient points, for each misconception we compile here the most direct, ``efficient'' and conclusive counterarguments:
\begin{itemize}[noitemsep,topsep=2pt,leftmargin=0.07\linewidth]
\item[(1)] `lines of force' misconception: field lines are characteristic of any vector field, not just the fields of force;
\item[(2)] `field lines as trajectories' misconception: not every vector field has a direct physical effect upon the particle motion; for the fields of force the field lines depend only on the field itself, while the trajectories depend also on the initial conditions;
\item[(3)] `field line density' misconception: field lines depend only on the vector field direction (as entailed even by their verbal description as tangential curves); a field magnitude may be varied at will, without any effect;
\item[(4)] `closed field lines' misconception: if a given type of vector field may vanish at any point---and the magnetic fields certainly can---then its field lines necessarily terminate at that point, hence not all can form closed loops.
\end{itemize}

We have had an opportunity to use some of these examples with students in class, through a free discussion within the time constraints imposed by the lectures. It was these discussions that allowed us to identify some of the students' objections that are presented in the paper (mostly felt in the sections about the field lines in connection to the force and the field lines as trajectories) and to clarify the counterarguments. Mostly, these objections were well-intended, a form of an honest intellectual curiosity (``what if we considered only a special type of field'' and such). We have a feeling that the gradual development and the clarification of ideas was satisfying and rewarding for them, as it certainly was for us.

Even if one were disposed to remove any mention of the field lines from educational practice, one could argue that `the damage has already been done'. Not only by the literature already brimming with them but also by the historical fact that the field lines \textit{were} a natural stepping stone toward the idea of a physical vector field. It would serve little purpose pretending that this could be rectified overnight. In all their glory, the field lines are rather nontrivial objects: they are solutions to a specific differential equation [Eq.~(\ref{master}) from Appendix~\ref{appendix_A}]. However, this line of presentation is rarely available to the instructors due to the mathematical capabilities of their target audience. As one can not always demonstrate everything that the field lines \textit{are}, at least show as soon as possible what they \textit{are not}. Explicitly expose students to the misconceptions that they are certain to encounter without our help; either as an assertion by some authoritative source (literature) or as an intuited conjecture based on their own extrapolations from an inadequate set of examples. Do so once in a timely and well guided manner, and build their intellectual immunity for the rest of time.

{

\appendix

\makeatletter
\@removefromreset{equation}{section}
\makeatother

\renewcommand{\theequation}{\arabic{equation}}

\section*{APPENDIX}

\vspace*{-1mm}

\section{For those who want more: field lines as trajectories} 
\label{appendix_A}

\vspace*{-1mm}

The general field line equation appears in literature in several different but equivalent forms~\cite{magnetic_1,electric_1,magnetic_3,magnetic_4,ajp_new,general_acc,plotting,rectilinear_acc}. Given $\mathbf{f}(\ell)$ as a vectorial field line parametrization by means of a parameter~$\ell$, the general field line equation for a vector field $\mathbf{F}(\mathbf{r})$ is most conveniently expressed as~\cite{our2}:
\begin{equation}
\frac{\D \mathbf{f}}{\D\ell}=\hat{\mathbf{F}}(\mathbf{f})=
\left\{\begin{array}{ccc}
\mathbf{F}(\mathbf{f})/|\mathbf{F}(\mathbf{f})| &\text{if}& \mathbf{F}(\mathbf{f})\neq\textbf{0},\\
\mathbf{0}&\text{if}& \mathbf{F}(\mathbf{f})=\textbf{0},
\end{array}\right.
\label{master}
\end{equation}
where a vector field $\hat{\mathbf{F}}$ of a \textit{unit} magnitude appears (with the exception of points where it vanishes, so its direction is not well defined). Solving this differential equation one obtains the field lines for any vector field $\mathbf{F}(\mathbf{r})$, without being limited to trivial or widely known examples. Most often this equation needs to be solved numerically, making this approach rather ungrateful for educational purposes. Many analytical solutions \textit{do} exist~\cite{electric_1,general_acc,rectilinear_acc,dipole,geo1,geo2,geo3,geo4}. However, they are not always easy to obtain, nor are they always elegant.

It is fortunate for the discussion from Section~\ref{trajectories} that there exists a simple (but not simple to derive) parametrization of the field lines from \fig~\ref{fig2}. Kristjansson~\cite{electric_1} obtains the parametrization for the `attractive dipole' (left plot), by using elliptic coordinates. Complementing his calculations by our own, we also find the parametrization for the `repulsive dipole' (right plot). Due to being remarkably similar, both cases may be written together in a compact form. In that, we use our own notation and express a planar solution from~\cite{electric_1} in cylindrical coordinates~$z$ and~$\rho$, thus obtaining a three-dimensional generalization. For two point charges \mbox{$|q_1|=|q_2|$} at \mbox{$\mathbf{r}_{1,2}=\pm L\,\hat{\mathbf{z}}$}, the particular-field-line coordinates may be parameterized as (a more general parametrization for \mbox{$|q_1|\neq|q_2|$} may be found in~\cite{ejp_new}):
\begin{align}
&z(t)=\pm Lt\sqrt{t\frac{2-\kappa t}{2t-\kappa}},
\label{zk_main}\\
&\rho(t)=\sigma L(1-t^2)\sqrt{\frac{\kappa}{2t-\kappa}}.
\label{rk_main}
\end{align}
The `attractive dipole' differs from the `repulsive' one by a sign~$\sigma$, by a geometric meaning of factor~$\kappa$ and by a range of a running parameter~$t$. The `attractive dipole' is characterized by \mbox{$\sigma=-1$}, the `repulsive' one by \mbox{$\sigma=+1$}. Thus it can be compactly expressed as \mbox{$\sigma=\mathrm{sgn}(q_1 q_2)$}, with $\mathrm{sgn}(\cdot)$ function returning a sign of its argument. Factor~$\kappa$ parameterizes a particular field line. It is related to the initial field line inclination (i.e. to a field line's spherical angle~$\theta_0$ as it leaves a charge at \mbox{$\mathbf{r}_1=L\,\hat{\mathbf{z}}$}) as \mbox{$\kappa=1+\sigma\cos\theta_0$}, so that \mbox{$\kappa\in[0,2]$} for both values of~$\sigma$. For the `attractive dipole' parameter~$t$ runs within a range \mbox{$t\in[1,2/\kappa]$}; for the `repulsive' one it runs within \mbox{$t\in\langle \kappa/2,1]$}. By using these equations one no longer needs to rely on drawing the field lines from \fig~\ref{fig2} by hand. Instead, one can present them exactly and correctly, without resorting to numerical solving of \eq~(\ref{master}).

As a sidenote to the `attractive dipole', a field line equation for a \textit{point dipole} is rather well known and much simpler to obtain~\cite{geo4}. For a point dipole of a dipole moment \mbox{$\mathbf{p}\propto\hat{\mathbf{z}}$} located at the origin, a field line solution in spherical coordinates reads:
\begin{equation}
r(\theta)=r_{\max}\sin^2\theta,
\label{point_dipole}
\end{equation}
where~$r_{\max}$ parameterizes a particular field line and corresponds to its maximal extent from the dipole itself. An exact solution may even be obtained for an oscillating point dipole of dipole moment \mbox{$\mathbf{p}(t)\propto\cos(\omega t)\hat{\mathbf{z}}$}~\cite{dipole}, and a static solution~(\ref{point_dipole}) may be recovered in the limit $\omega\to0$. In fact, a solution in spherical coordinates is known for a static point multipole of an arbitrary degree $n$ (dipole $n=1$, quadrupole $n=2$, ...)~\cite{geo1,geo2}:
\begin{equation}
r(\theta)=r_0 \left| \sin\theta \, P_n^1(\cos\theta) \right|^{1/n},
\label{point_multipole}
\end{equation}
with associated Legendre polynomial $P_n^m(\cdot)$, and~$r_0$ as an integration constant specifying a particular field line. Due to a requirement \mbox{$n\ge m$} and a specific value of \mbox{$m=1$} appearing in \eq~(\ref{point_multipole}), this parametrization evidently does not hold for a charged monopole (\mbox{$n=0$}). On the other hand, a dipole solution can immediately be confirmed to reproduce a well known case~(\ref{point_dipole}), since \mbox{$P_1^1(\cos\theta)=-\sin\theta$}.

\subsection*{Trajectories can intersect, field lines can not}

We dedicate this section to another incompatible property between the field lines and particle trajectories. \textit{While the field lines can not intersect (neither themselves nor any other field line\footnote{
If the field lines or single-line's segments intersected, the vector field at the intersection point should simultaneously point in different directions, along each of the locally crossed segments.
}), the trajectories can}. 

This is a sufficient proof that they can not categorically coincide. In principle, one only needs to identify one example of \textit{different} trajectories crossing to conclude a proof by counterexample. Consider then the trajectories of two balls in a homogenous gravitational field thrown toward some common point, so that their paths do indeed intersect, not necessarily simultaneously.

It would also be instructive to identify a clear example of a single trajectory crossing \textit{itself}. To this end we will call upon some well known results from the classical mechanics, in order to show how much of a `common knowledge' one can use to refute the field-lines-as-trajectories misconception. The first result is known as Bertrand's theorem~\cite{bertrand_1,bertrand_2}. It states that among all central fields of force---the fields of form \mbox{$\mathbf{F}(\mathbf{r})=F(r)\hat{\mathbf{r}}$}---only in the familiar inverse-square field \mbox{$F(r)\propto 1/r^2$} and the linear ``elastic'' field \mbox{$F(r)\propto r$} are \textit{all} bounded trajectories closed. This claim is often encountered in connection to a remarkable fact that all bounded trajectories in a classical point-mass gravitational field are ellipses. Its converse---that among bounded trajectories (if such exist at all) in any other central field of force some are necessarily open---strongly suggests that we might find some self-intersecting trajectories (being constrained to a finite portion of space, while not being able to close upon themselves). Thus, Bertrand's theorem motivates us go in search of some such example. For instructors' convenience we propose here a field of force from the so-called Manev problem~\cite{manev}:
\begin{equation}
\mathbf{F}(\mathbf{r})=-\left(\frac{A}{r^2}+\frac{B}{r^3}\right)\hat{\mathbf{r}},
\label{manev}
\end{equation}
for the simple reason that the resulting trajectories have a simple analytical solution. Being constrained to a plane due to a conservation of angular momentum, they are best expressed in polar coordinates ($\rho,\varphi$), parametrizing a given trajectory as \mbox{$\mathbf{r}=\rho\hat{\boldsymbol{\rho}}$}. For those $A,B$ that can yield bounded trajectories, the bounded solutions may be expressed as:
\begin{equation}
\rho(\varphi)=\frac{p}{1+k\cos[\Omega(\varphi-\varphi_0)]},
\label{manev_sol}
\end{equation}
with parameters \mbox{$p,k,\Omega,\varphi_0$} deepening on the initial conditions. Playing with these parameters---crucially, with~$k$ and~$\Omega$---one can easily identify a plethora of self-intersecting trajectories. Many are open\footnote{
Or at least \textit{seem} open. Since their openness depends on~$\Omega$ being irrational, on a personal computer all the trajectories will close upon themselves after a sufficient number of revolutions, since every computer-representable number is rational.
}, in accordance with Bertrand's theorem, but some are closed (those for rational values of~$\Omega$). Figure~\ref{figx} presents an example of a closed trajectory intersecting itself (the search for it having being motivated by a theorem that guarantees the existence of open trajectories), obtained with \mbox{$k=0.5$} and \mbox{$\Omega=1.5$}, in arbitrary units.\\

\begin{figure}[t!]
\centering
\includegraphics[height=\figheight\textwidth,keepaspectratio]{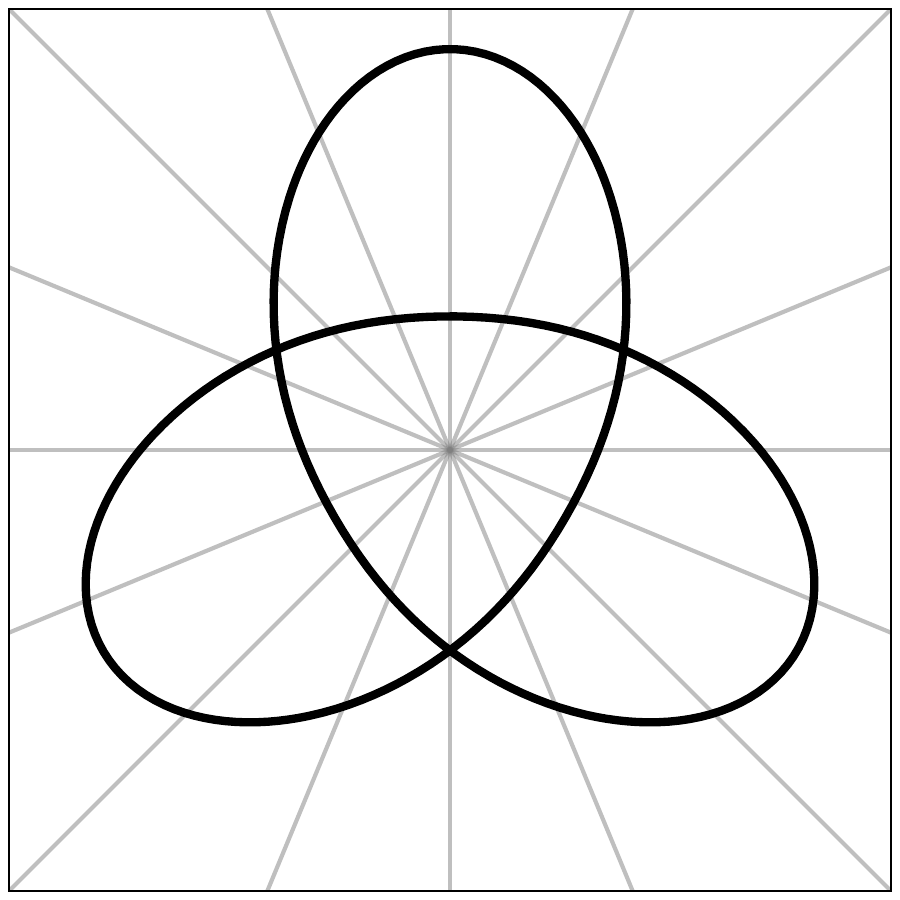}
\caption{Another counterexample to the field-lines-as-trajectories misconception. Thick black line shows a self-intersecting closed trajectory from \eq~(\ref{manev_sol}) for \mbox{$k=0.5$} and \mbox{$\Omega=1.5$}. Thin grey lines show the field lines of an associated field of force from \eq~(\ref{manev}).}
\label{figx}
\end{figure}

Non-trivial self-intersecting trajectories, such as the one from \fig~\ref{figx}, can only be obtained with the initial velocity having a non-vanishing azimuthal component. In other words, they can not be obtained by letting a particle into motion from rest (in which case the trajectories fully lie along the field lines). Though this in no way invalidates or even weakens a disproof of the field-lines-as-trajectories misconception, it would also be handy for instructors to have at the ready a counterexample with a self-intersecting trajectory staring from rest. This would be analogous to \fig~\ref{fig2} from the main text satisfying the students' possible insistence on the particle's initial rest.

To this end we consider another widely familiar example: a two-dimensional anisotropic harmonic oscillator, i.e. an associated field of force:
\begin{equation}
\mathbf{F}(\boldsymbol{\rho})=-k_x \mathbf{x}-k_y \mathbf{y},
\label{ani_ho}
\end{equation}
with \mbox{$k_x\neq k_y$}. A general particle trajectory corresponds to the so-called Lissajous curve~\cite{lissajous}, parametrised as: \mbox{$x(t)=A_x\cos(\omega_x t-\delta_x)$} and \mbox{$y(t)=A_y\cos(\omega_y t-\delta_y)$}---with \mbox{$\omega_{x,y}=\sqrt{k_{x,y}/m}$} and~$m$ the particle mass---wherein the integration constants~$A_{x,y},\delta_{x,y}$ follow from the initial conditions. For a particle initially \textit{at rest}, let into motion from the initial position \mbox{$(x_0,y_0)$}, the solution reduces to:
\begin{equation}
x(t)=x_0\cos(\omega_x t) \quad\text{and}\quad y(t)=y_0\cos(\omega_y t),
\label{lissajous}
\end{equation}
fully determining a given Lissajous-curve trajectory.

\begin{figure}[t!]
\centering
\includegraphics[height=\figheight\textwidth,keepaspectratio]{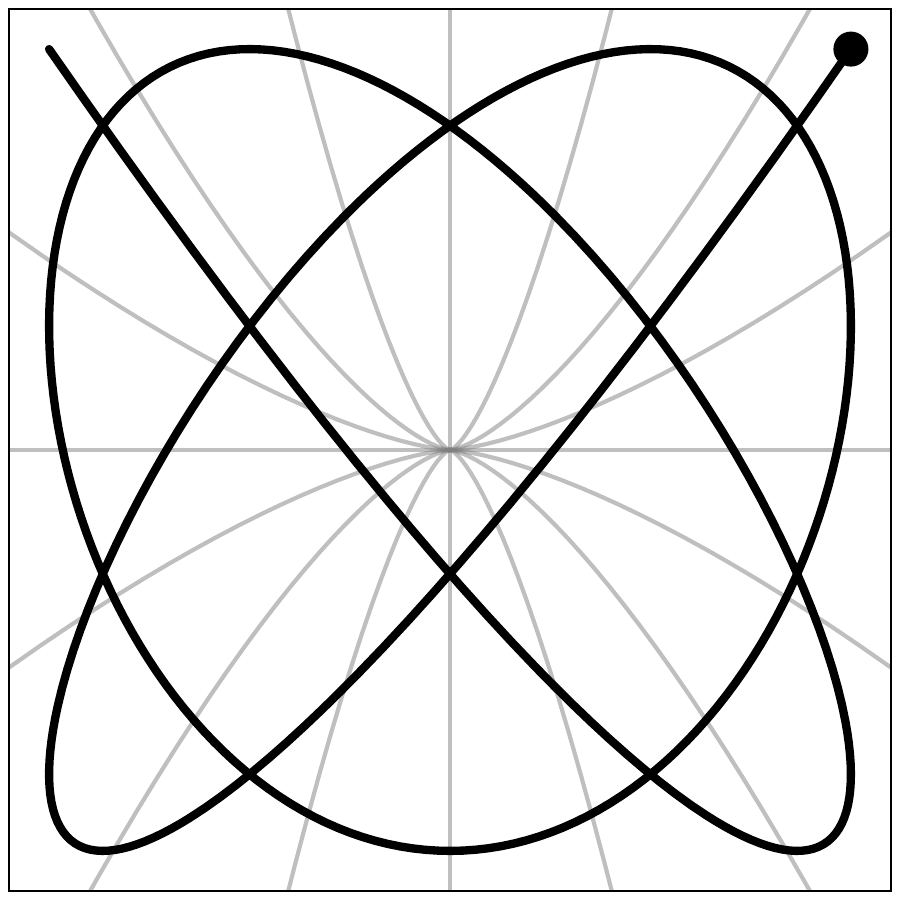}
\caption{Yet another counterexample to the field-lines-as-trajectories misconception. Thick black line shows a self-intersecting Lissajous-curve trajectory from \eq~(\ref{lissajous}), for \mbox{$x_0=y_0$} and \mbox{$\omega_y/\omega_x=1.2$}. Thin grey lines show the field lines from \eq~(\ref{ani_ho_line}), with \mbox{$k_y/k_x=(\omega_y/\omega_x)^2$}. The trajectory starts \textit{from rest}, from the top right corner (the black point).}
\label{figy}
\end{figure}

If we now found a single self-intersecting trajectory by trying out different combinations of $\omega_x$ and $\omega_y$, our work of disproving the field-lines-as-trajectories misconception would be complete. But for a clear comparison beyond any shadow of a doubt, let us also determine the exact form of the field lines for a field from \eq~(\ref{ani_ho}). To this end we select the Cartesian coordinates in parametrizing the field line increments \mbox{$\D\mathbf{f}=\D x\,\hat{\mathbf{x}}+\D y\,\hat{\mathbf{y}}$}, reflecting the field parametrization \mbox{$\mathbf{F}=F_x\hat{\mathbf{x}}+F_y\hat{\mathbf{y}}$}, with \mbox{$F_x=-k_x x$} and \mbox{$F_y=-k_y y$}. The general field line equation~(\ref{master}) provides an operational differential equation:
\begin{equation}
\frac{\D x}{F_x}=\frac{\D y}{F_y} \quad\Rightarrow\quad \frac{\D y}{\D x}=\frac{F_y}{F_x}=\frac{k_y}{k_x}\frac{y}{x}.
\end{equation}
Its solutions are easily found by a direct integration:
\begin{equation}
y(x)=Y_0\Big(\frac{x}{X_0}\Big)^{k_y/k_x},
\label{ani_ho_line}
\end{equation}
with \mbox{$(X_0,Y_0)$} as a desired point through which a given field line passes. Figure~\ref{figy} shows one self-intersecting Lissajous-curve trajectory from \eq~(\ref{lissajous})---for \mbox{$x_0=y_0$} and \mbox{$\omega_y/\omega_x=1.2$}---comparing it against the field lines from \eq~(\ref{ani_ho_line}), obtained using the associated exponent \mbox{$k_y/k_x=(\omega_y/\omega_x)^2$}.

\section{For those who want more: field lines density as a measure of a field magnitude}
\label{appendix_B}

\begin{table*}[t!]
\caption{\label{tab1} Examples of field component ratios $F_\rho/F_\varphi$ from \eq~(\ref{diff_form}), yielding different field line density dependences on~$\rho$. Corresponding field lines are shown in \fig~\ref{fig5}. In each case three different field line magnitude dependences on~$\rho$ are constructed. Magnitude dependences inconsistent with density dependences are marked in bold font.}
\vspace*{4mm}
\begin{tabular}{lclccclcc}
\hline\hline
\vtop{\hbox{\strut Ratio of field}\hbox{\strut components}} &&
\vtop{\hbox{\strut Field line}\hbox{\strut equation}} &&
\vtop{\hbox{\strut Field line density}\hbox{\strut dependence on $\rho$}} &&
\vtop{\hbox{\strut Field}\hbox{\strut example}}&&
\vtop{\hbox{\strut Field magnitude}\hbox{\strut dependence on $\rho$}} \\
\hline
\multirow{3}{*}{$\displaystyle \frac{F_\rho}{F_\varphi}=\frac{\varepsilon}{\mathcal{E}}$} && \multirow{3}{*}{$\displaystyle \rho=\rho_0\exp\left[\frac{\varepsilon}{\mathcal{E}}(\varphi-\varphi_0)\right]$} && \multirow{3}{*}{decreasing} && $\mathbf{F}=\rho^{-1}\left(\varepsilon\,\hat{\boldsymbol{\rho}}+\mathcal{E}\,\hat{\boldsymbol{\varphi}}\right)$ && decreasing \\
&& && && $\mathbf{F}=\varepsilon\,\hat{\boldsymbol{\rho}}+\mathcal{E}\,\hat{\boldsymbol{\varphi}}$ && \textbf{constant} \\
&& && && $\mathbf{F}=\rho\left(\varepsilon\,\hat{\boldsymbol{\rho}}+\mathcal{E}\,\hat{\boldsymbol{\varphi}}\right)$ && \textbf{increasing} \\
\hline
\multirow{3}{*}{$\displaystyle \frac{F_\rho}{F_\varphi}=\frac{\varepsilon}{\mathcal{E}}\rho^{-1}$} && \multirow{3}{*}{$\displaystyle \rho=\rho_0+\frac{\varepsilon}{\mathcal{E}}(\varphi-\varphi_0)$} && \multirow{3}{*}{constant} && $\mathbf{F}=\rho^{-1}\left(\varepsilon\,\hat{\boldsymbol{\rho}}+\mathcal{E}\rho\,\hat{\boldsymbol{\varphi}}\right)$ && \textbf{decreasing} \\
&& && && $\mathbf{F}=\frac{1}{\sqrt{\varepsilon^2+(\mathcal{E}\rho)^2}}\left(\varepsilon\,\hat{\boldsymbol{\rho}}+\mathcal{E}\rho\,\hat{\boldsymbol{\varphi}}\right)$ && constant \\
&& && && $\mathbf{F}=\varepsilon\,\hat{\boldsymbol{\rho}}+\mathcal{E}\rho\,\hat{\boldsymbol{\varphi}}$ && \textbf{increasing} \\
\hline
\multirow{3}{*}{$\displaystyle \frac{F_\rho}{F_\varphi}=\frac{\varepsilon}{\mathcal{E}}\rho^{-2}$} && \multirow{3}{*}{$\displaystyle \rho=\sqrt{\rho_0^{2}+\frac{2\varepsilon}{\mathcal{E}}(\varphi-\varphi_0)}$} && \multirow{3}{*}{increasing} && $\mathbf{F}=\rho^{-2}\left(\varepsilon\,\hat{\boldsymbol{\rho}}+\mathcal{E}\rho^2\,\hat{\boldsymbol{\varphi}}\right)$ && \textbf{decreasing} \\
&& && && $\mathbf{F}=\frac{1}{\sqrt{\varepsilon^2+(\mathcal{E}\rho^2)^2}}\left(\varepsilon\,\hat{\boldsymbol{\rho}}+\mathcal{E}\rho^2\,\hat{\boldsymbol{\varphi}}\right)$ && \textbf{constant} \\
&& && && $\mathbf{F}=\varepsilon\,\hat{\boldsymbol{\rho}}+\mathcal{E}\rho^2\,\hat{\boldsymbol{\varphi}}$ && increasing \\
\hline\hline
\end{tabular}
\end{table*}

\begin{figure*}[t!]
\centering
\includegraphics[height=\figheight\textwidth,keepaspectratio]{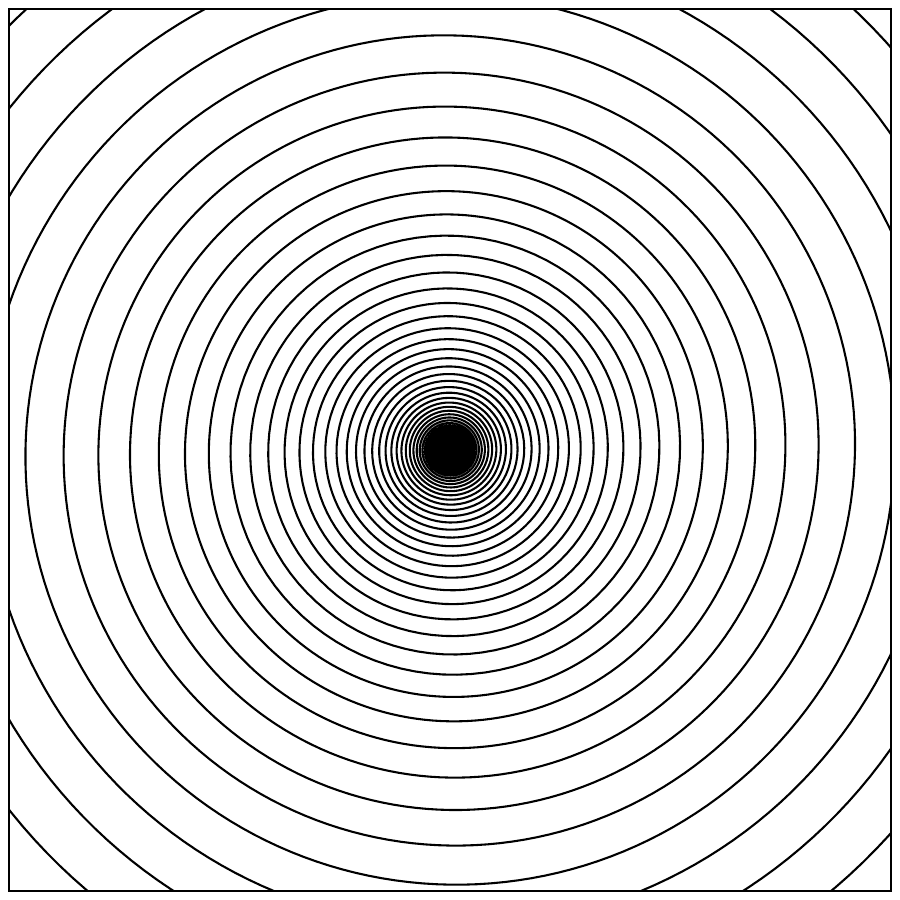}
\includegraphics[height=\figheight\textwidth,keepaspectratio]{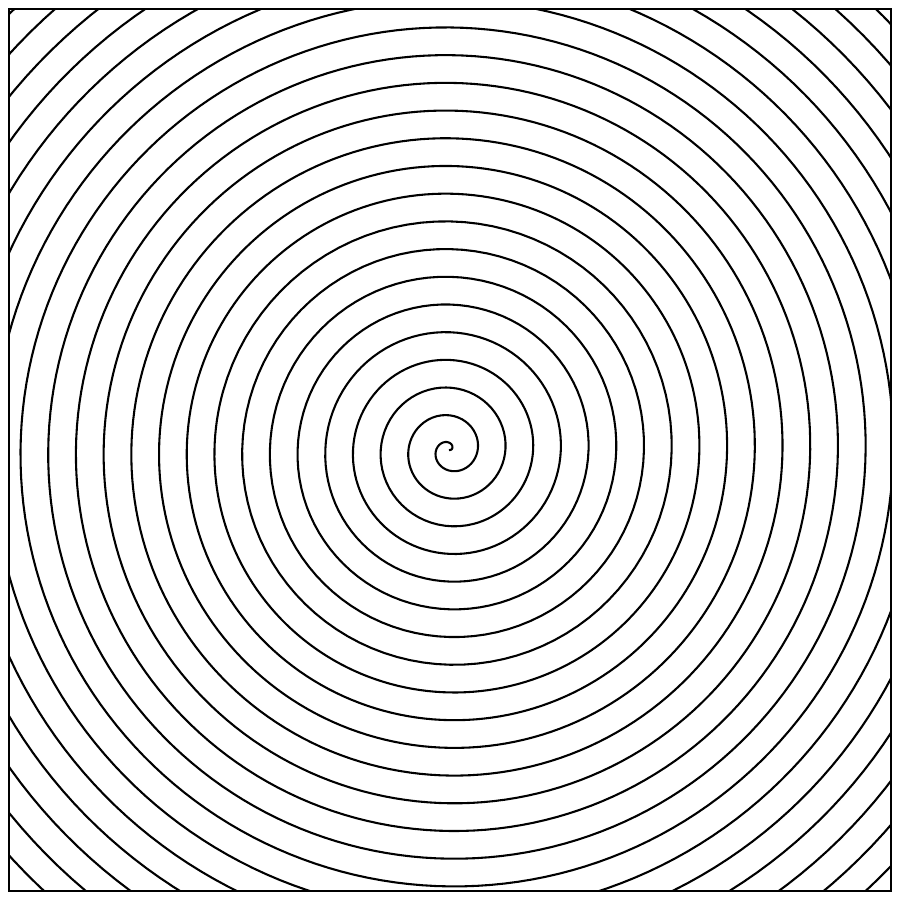}
\includegraphics[height=\figheight\textwidth,keepaspectratio]{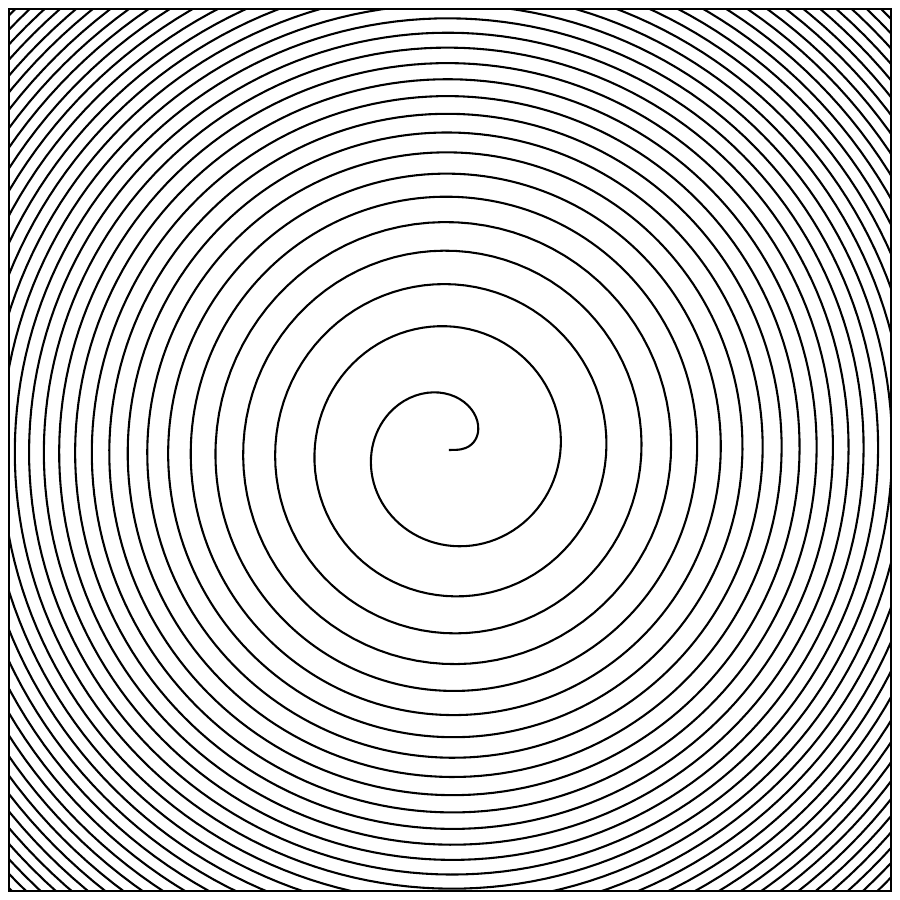}
\caption{Field line examples from table~\ref{tab1} (field lines listed therein from top to bottom are shown here from left to right).
}
\label{fig5}
\end{figure*}

In order to calculate the particular field lines of a field~(\ref{E_spiral}) from the general field line equation~(\ref{master}), we now select the polar coordinates, thus parametrizing: \mbox{$\D\mathbf{f}=\D\rho\,\hat{\boldsymbol{\rho}}+\rho\,\D\varphi\,\hat{\boldsymbol{\varphi}}$} and \mbox{$\mathbf{F}=F_\rho\hat{\boldsymbol{\rho}}+F_\varphi\hat{\boldsymbol{\varphi}}$}. With this selection \eq~(\ref{master}) may be recast as:
\begin{equation}
\frac{\D \rho}{F_\rho}=\frac{\rho\,\D\varphi}{F_\varphi} \quad\Rightarrow\quad \frac{1}{\rho}\frac{\D\rho}{\D\varphi}=\frac{F_\rho}{F_\varphi},
\label{diff_form}
\end{equation}
which again gives us a particular differential equation to be solved. It is also obvious from this form---though less so than from \eq~(\ref{master})---that a field magnitude does not enter a field line determination, as any rescaling of $\mathbf{F}(\mathbf{r})$ by a (strictly positive) scalar function $f(\mathbf{r})$ leaves the ratio \mbox{$F_\rho/F_\varphi$} unaffected. Since the field lines are completely determined by a ratio of the field components, we may do what we will with its magnitude, as long as we keep its direction unchanged. That means that by varying both a direction and a magnitude of a field $\mathbf{F}(\mathbf{r})$, we can manipulate its field lines and its magnitude \textit{independently}.

Table~\ref{tab1} lists simple examples of these independent manipulations. Three behaviors for a ratio \mbox{$F_\rho/F_\varphi$} are presented, yielding the different field line density dependences on~$\rho$, which are shown in \fig~\ref{fig5}. Field line equations are given in each case, obtained by solving \eq~(\ref{diff_form}). In that, $\rho_0$ and $\varphi_0$ are the initial coordinates for tracing out a particular field line. In each case three particular field realizations are given, each with a different magnitude dependence on~$\rho$. By adjusting the field lines independently from a field magnitude, one easily obtains inconsistent behaviors marked by a bold font. A trivial manner in which these field realizations were constructed should be appreciated, as for a given ratio \mbox{$F_\rho/F_\varphi$} we achieve a desired magnitude dependence by an arbitrary selection of a scalar prefactor. Let us summarize the observations (including those from Section~\ref{density}) disproving a relation between a field line density and a field magnitude: (1)~in all these examples they follow an opposite dependence on~$\varepsilon$: as density increases, magnitude decreases and vice versa; (2)~as $\varepsilon$ approaches 0, density diverges, while magnitude remains finite; (3)~their dependence on~$\rho$ can be manipulated independently and can easily be selected as inconsistent.

}

\onecolumngrid



\clearpage


\setcounter{page}{1}
\setcounter{section}{0}
\setcounter{footnote}{0}

\renewcommand{\thesection}{\Alph{section}}
\numberwithin{equation}{section}
\numberwithin{figure}{section}
\numberwithin{table}{section}

\renewcommand{\theequation}{\thesection\arabic{equation}}
\renewcommand{\thefigure}{\thesection\arabic{figure}}

\renewcommand{\thepage}{S\arabic{page}}  

\settopmargin{1.9cm} 
\setbottommargin{2.5cm} 

\raggedbottom

\onecolumngrid
\begin{center}

\textbf{\LARGE Supplementary note}\\[.5cm]

\textbf{\large A stop to field line misconceptions}\\[.5cm]

Petar \v{Z}ugec$^{1}$, Ivica Fri\v{s}\v{c}i\'{c}$^{1}$, Mihael Makek$^{1}$, Eric Andreas Vivoda$^{1}$\\[.1cm]
{\small
{\itshape
$^1$University of Zagreb Faculty of Science, Department of Physics, Zagreb, Croatia\\}
}


\end{center}


\begin{center}
\begin{minipage}{400pt} 
\small
Of all misconceptions addressed in the main paper, the one about a field line density (Section~IV) is the most difficult to dispel. For this reason we compile here some further examples illustrating a general failure of two-dimensional field line diagrams in conveying information about the (electric) field magnitude. These examples are somewhat more technical, in a sense that either no closed-form solution for the field lines seems to exist---so that one must resort to numerical solving of the differential field line equation~(\ref{master}) from the main paper---or we attach an involved technical calculation alongside a given example.\\
\end{minipage}
\end{center}

\twocolumngrid

\section{False monopole moment}

\begin{figure}[t!]
\centering
\includegraphics[height=0.3\textwidth,keepaspectratio]{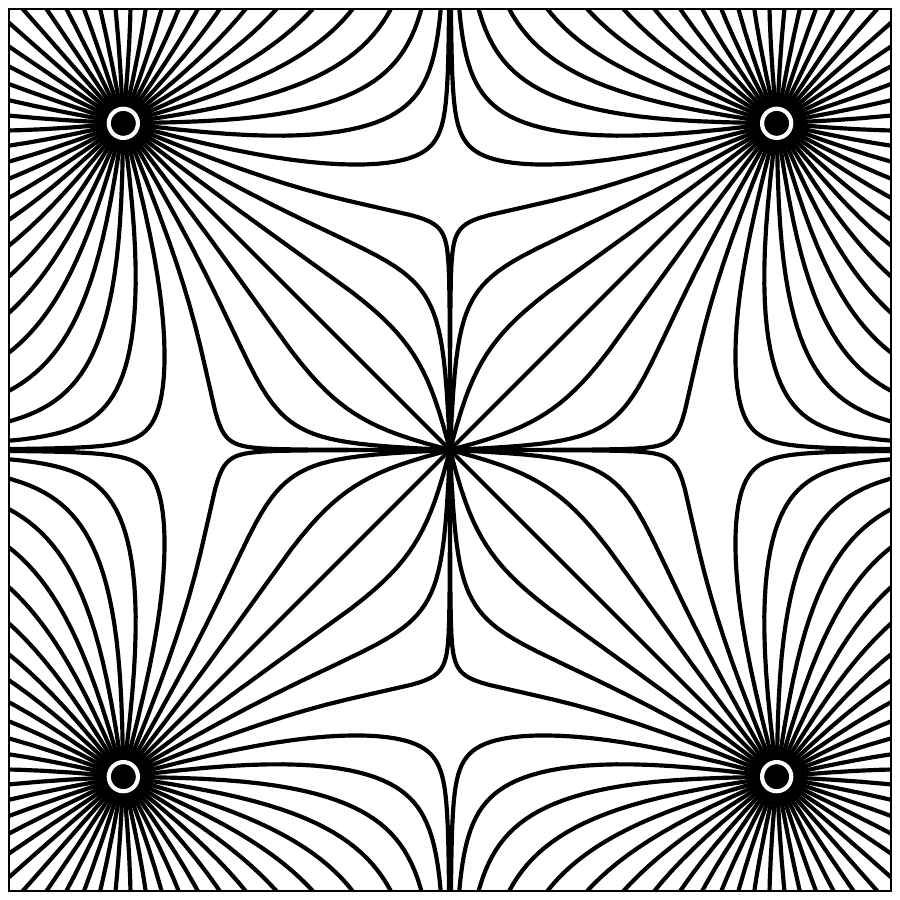}
\caption{Field lines due to four equal charges placed in a square configuration. There is no charge at the center. The field lines not only suggest an existence of a false monopole, but their density diverges where a field magnitude vanishes.}
\label{Sfig1}
\vspace*{2mm}
\end{figure}

The following beautiful example by Freeman~\cite{false_monopole} illustrates in a painfully obvious way the extreme dangers in appraising the electric field magnitude from planar (two-dimensional) field line diagrams. Consider four equal charges, each placed at a corner of a square. We are not aware of any closed solution for the field lines of this configuration, so we obtain them as a numerical solution to a differential field line equation~(\ref{master}) from the main paper. Figure~\ref{Sfig1} shows the results. Notice that there is no charge present at the origin, i.e. at the center of a square configuration! Yet, within this particular plane containing the four source charges, an entire family of field lines converges to a central point devoid of any charge, suggesting a presence of a \textit{false monopole}.

But not even this is of central interest to us. Rather, we are interested in the relation between the field at the origin and a density of field lines around it. From the symmetry of the setup it is perfectly clear that the electric field at the origin vanishes due to a vectorial cancellation. On the other hand, the convergence of a continuous set of field lines to this single point makes a planar field line density divergent! We are referring to a density that visually imposes itself, which is a density $\varrho$ per arc length $\D\ell=\rho\,\D\varphi$ around the origin:

\begin{equation}
\varrho(\varphi)\equiv\frac{\D N}{\D \ell}=\frac{1}{\rho}\frac{\D N}{\D\varphi},
\end{equation}
at a radial distance~$\rho$ from it. (See the next section for a detailed elaboration of the field lines density definition, in particular a discussion regarding~$\D N$ as a measure of the amount of field lines.) This density diverges as \mbox{$\rho\to0$}, while the amount of field lines~$\D N$ approaching the origin \textit{does not decrease in proportion to~$\rho$}. Notice that this is the same type of density that is supposed to convey a radial dependence and the divergent nature of an electrostatic field of a \textit{single} point charge; a density that diverges through the same mathematical mechanism as described here. The only difference is that the density here is anisotropic. And that there is no actual charge.

\section{Linear quadrupole}

\begin{figure}[t!]
\centering
\includegraphics[height=0.3\textwidth,keepaspectratio]{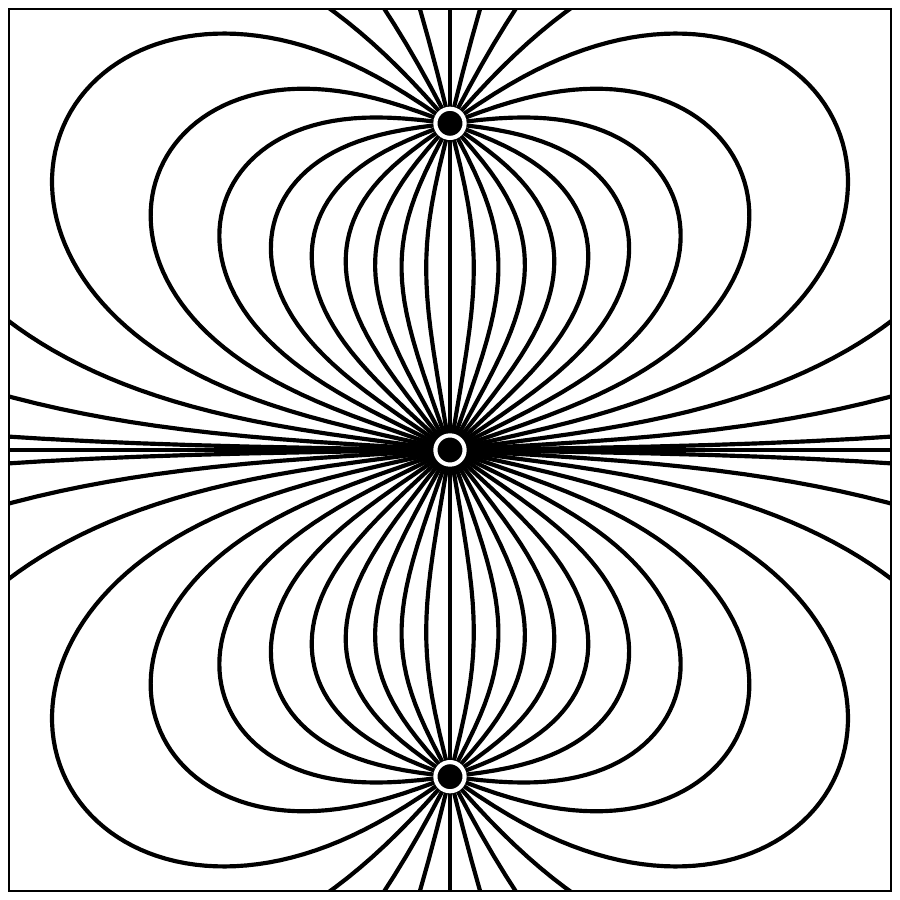}
\caption{Linear quadrupole consisting of two outer charges~$q_0$ and a single charge~$-2q_0$ exactly half-way between them. The field lines leave the outer charges isotropically (in equal angular steps) within this particular plane.}
\label{Sfig2}
\end{figure}

This example is one of the more elegant ones by Wolf et al.~\cite{quadrupole}. It consists of a linear quadrupole, wherein a single charge of magnitude~$-2q_0$ is placed exactly half-way between the two charges of magnitude~$q_0$. A configuration is shown in \fig~\ref{Sfig2}, together with the numerically obtained field lines which have been generated in equal angular steps around the two outer charges and have been numerically propagated toward the central charge (due to a specific 2:1 ratio between the central and outer charges, \textit{all} field lines---save for the obvious two---connect to a central charge). The \textit{planar isotropy} of the selected field lines around the outer charges represents the isotropy of the electrostatic field in their immediate vicinity. If the field line density continued to properly represent a field magnitude within this plane, then the field lines should again be uniformly distributed around the central charge. This isotropy would, of course, represent an isotropy of electrostatic field in the immediate vicinity of the central charge, where its own field dominates over any contribution from the outer charges, due to the familiar \mbox{$1/r^2$} dependence. However, it is clearly visible that this isotropy is not realized and---if we took the outer charges to be at the poles---the field lines ending at the central charge increasingly gather around its equator. This effect is called \textit{equatorial clumping}~\cite{quadrupole}. In the general case it simply makes it impossible to draw any consistent conclusions about the field magnitude from two-dimensional field line diagrams, whether the field be electrostatic or otherwise.

\begin{figure}[t!]
\centering
\includegraphics[width=1\linewidth,keepaspectratio]{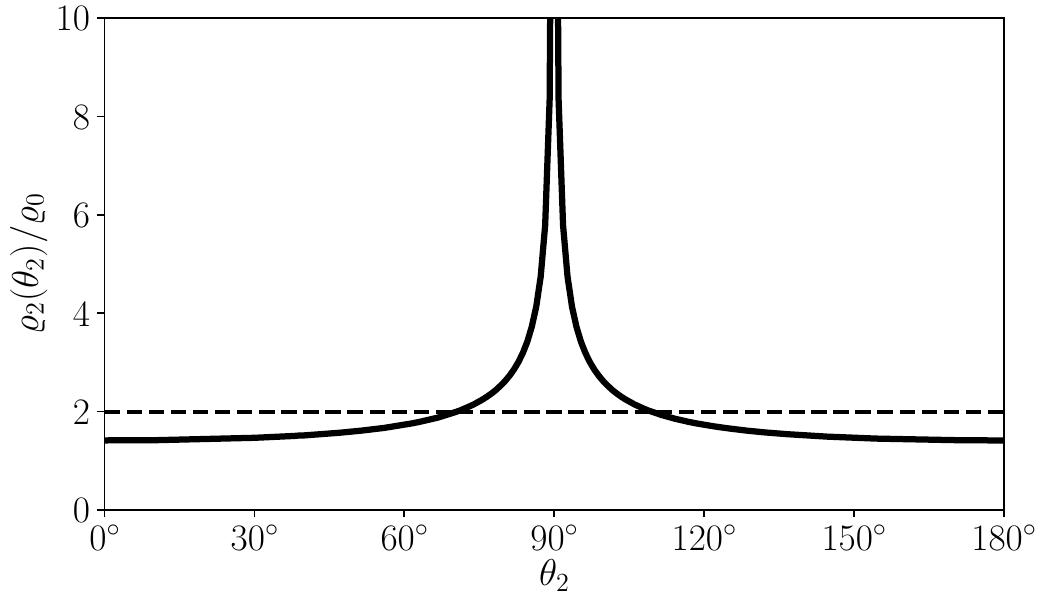}
\caption{Relative angular density of the field lines around the central charge from \fig~\ref{Sfig2}, obtained numerically from $10^5$ field lines generated in uniform angular steps around each outer charge. It is in perfect agreement with an analytical form~(\ref{quad_analytic}), demonstrating an integrable singularity.} 
\label{Sfig3}
\end{figure}

In the absence of the closed-form field line solutions, one can easily analyze their angular density around the central charge by a numerical procedure. First generate a multitude of initially-2D-isotropic starting points in the immediate vicinity of outer charges. By numerically solving the differential field line equation~(\ref{master}) from the main paper for each of these staring points, propagate the field lines until they reach the central charge. Their angle of approach to the central charge is now easily extracted from each numerical field line solution and a distribution of these angles can be plotted. From \fig~\ref{Sfig3} one can immediately glance the results for $10^5$ field lines uniformly generated around each outer charge. But let us first define the circumstances and parameters required to put the figure into perspective.

For simplicity we consider only a portion of space above the central charge (housing the upper outer charge), and use the symmetry of the setup to recover the results from below the central charge (for the lower outer charge). Let the $z$-axis of the coordinate system pass vertically upwards through all the charges. Let~$\theta_1$ be the initial field line angle as it leaves the upper charge; let~$\theta_2$ be the terminating field line angle as it reaches the central charge. Let both angles be defined relative to the positive direction of the $z$-axis\footnote{Both~$\theta_1$ and~$\theta_2$ correspond to the typically defined spherical angles. Accordingly, both angles will cover an angular range \mbox{$[0,\pi]$}. We use them for a 2D case in order to easily achieve a 3D generalization at a later point. Thus defined, they cover only a half of the available 2D space, while the results from the other half are easily recovered from a left-right symmetry of the setup.
}. Let us define an \textit{angular density} \mbox{$\varrho_i(\theta_i)$} of the field lines around the outer ($i=1$) or the central ($i=2$) charge as:
\begin{equation}
\varrho_i(\theta_i)\equiv\frac{\D N}{\D \theta_i},
\label{ang_dens}
\end{equation}
with~$\D N$ as a measure of the number of field lines either leaving the outer charge or approaching the central charge within a corresponding angular interval~$\D\theta_i$. When dealing with the numerically generated field lines, their total number~$N$ will, of course, be finite. But we will soon treat the definition~(\ref{ang_dens}) analytically, whereby we will consider the distribution of field lines as continuous (i.e. the field lines as continuously varying from one to the other, filling the entire space). We will call such set of field lines as \textit{continual field lines}, as opposed to a discrete set of the numerically obtained ones. In case of the continual field lines~$\D N$ is to be treated as a \textit{measure} of their quantity, as even their differential amount within~$\D\theta_i$ is infinite. This creates no logical difficulties as one can meaningfully speak only of the \textit{relative} densities, such as \mbox{$\varrho_2/\varrho_1$}. Otherwise, one can obtain consistent results simply by pronouncing a measure~$N_0$ of the total number of continual field lines to be some arbitrary number (e.g. \mbox{$N_0=1$}), thus imposing a density normalization: \mbox{$\int_0^\pi \varrho(\theta)\D\theta=N_0$}.
$$\ast\ast\ast$$

\vspace*{2mm}

We return now to the $10^5$ numerically generated field lines from each outer charge. A key point is that they be generated as leaving the upper charge with a uniform angular density:
\begin{equation}
\varrho_1(\theta_1)=\varrho_0,
\label{varrho0}
\end{equation}
$\varrho_0$ being a fixed, constant value. Numerically extracting a distribution of terminating angles at the central charge, one obtains a relative density \mbox{$\varrho_2(\theta_2)/\varrho_0$} from \fig~\ref{Sfig3}, wherein a sharp peak represents the equatorial clumping effect. The dashed line is a mean value~\mbox{$\bar{\varrho}_2/\varrho_0=2$}\linebreak

\noindent of a relative distribution\footnote{
Factor 2 relative to $\varrho_0$ comes from an angular compression of the field liens around the central charge. The field lines leaving the upper charge, and covering an angular range \mbox{$\theta_1\in[0,\pi]$} around it, terminate within a range \mbox{$\theta_2\in[0,\pi/2\rangle$} around the central charge. A normalization \mbox{$\int_0^\pi \varrho_0\D\theta_1=\int_0^{\pi/2} \bar{\varrho}_2\D\theta_2$} requires \mbox{$\bar{\varrho}_2/\varrho_0=2$.} 
}. It shows how the density of the terminating angles would look if they were uniformly distributed around the central charge, i.e. if the planar field line density accurately represented a field magnitude there. We not only observe that the density is not uniform; rather, it looks like it might be divergent! And indeed it is, which can be shown analytically.

To this end, we consider a `conservation of field lines'. Since all field lines that start at the upper charge terminate at the central charge, their measure~$\D N$ from~(\ref{ang_dens}) must be the same at both charges, leading to:
\begin{equation}
\varrho_1(\theta_1)|\D\theta_1|=\varrho_2(\theta_2)|\D\theta_2|.
\label{conser}
\end{equation}
Absolute values appear because $\theta_1$ and $\theta_2$, when mutually dependent, may vary in different directions, i.e. their increments $\D\theta_1$ and $\D\theta_2$ may have opposite signs. This immediately gives us a general transformation rule:
\begin{equation}
\frac{\varrho_2(\theta_2)}{\varrho_1(\theta_1)}=\left|\frac{\D\theta_1}{\D\theta_2}\right|
\label{transfor}
\end{equation}
for calculating the relative field line densities. Evidently, we have to find the relation between $\theta_1$ and $\theta_2$, regarding each particular field line. Lacking the closed-form field line solutions, we follow the procedure by Wolf et al.~\cite{quadrupole}. Based on the Gauss's law\footnote{
These are the  same considerations that lead to a relation~(\ref{rho_relation}) from the main paper, which is a correct relation between a field line density and a field magnitude for divergence-free fields. In that, all applications of the Gauss's law must be carried out in a three-dimensional space, since for $1/r^2$ fields it only holds in 3D. 
} they first consider an electric flux from the upper charge, contained within the \textit{outgoing endcap} with an angular opening $\theta_1$; an endcap bounded by all the field lines leaving the same charge under~$\theta_1$. All these field lines terminate at the central charge under an angle~$\theta_2$. In that, all field lines from within the outgoing endcap (for \mbox{$\theta\in[0,\theta_1]$}) terminate between~$\theta_2$ and the equator (under \mbox{$\theta\in[\theta_2,\pi/2\rangle$}). Following Wolf et al., we will call this angular range a \textit{terminating ring}. Since, by definition, a field is everywhere tangential to the field lines, no electric flux can `cross over' them and the total flux from the outgoing endcap stays contained within the terminating ring. In the immediate vicinity of each charge (at a distance \mbox{$r\to0$}), where their fields are radial and dominated by a given charge, the flux within an outgoing cap is \mbox{$E_1r^2\Omega_1$}, and the flux within a terminating ring is \mbox{$E_2r^2\Omega_2$}; $E_1$ and $E_2$ being the field magnitudes, with $\Omega_1$ and $\Omega_2$ as the subtended solid angles. Taking into account that the central charge \mbox{$q_2=-2q_0$} has double the magnitude of the outer charge \mbox{$q_1=q_0$}, their charge-immediate fields satisfy: $E_2/E_1=|q_2/q_1|=2$. Thus, equating the earlier fluxes leads to:
\begin{equation}
E_1r^2\Omega_1=E_2r^2\Omega_2 \quad\Rightarrow\quad \Omega_2=\Omega_1/2.
\label{solid}
\end{equation}

\vspace*{3.7mm}

\pagebreak

\noindent It is now a simple matter calculating the solid angles:
\begin{align}
& \Omega_1=2\pi\int_0^{\theta_1} \sin\theta\,\D\theta=2\pi(1-\cos\theta_1),
\label{solid1}\\
&\Omega_2=2\pi\int_{\theta_2}^{\pi/2} \sin\theta\,\D\theta=2\pi \cos\theta_2.
\label{solid2}
\end{align}
Plugging these results into~(\ref{solid}) yields the sought relation between $\theta_1$ and $\theta_2$:
\begin{equation}
\theta_1=\arccos(1-2\cos\theta_2).
\label{th_1}
\end{equation}
Differentiating this, as required by~(\ref{transfor}), generalizing the result for the lower outer charge (i.e. for terminating angles \mbox{$\theta_2\in\langle \pi/2,\pi]$}) and using~(\ref{varrho0}) gives us an analytical form of the relative density from \fig~\ref{Sfig3}:
\begin{equation}
\frac{\varrho_2(\theta_2)}{\varrho_0}=\frac{\sin\theta_2}{\sqrt{(1-|\cos\theta_2|)\,|\cos\theta_2|}}, 
\label{quad_analytic}
\end{equation}
which we find to be in perfect agreement with the numerically extracted one. It can immediately be seen from~(\ref{quad_analytic}) that the relative density is indeed divergent for $\theta_2\to\pi/2$, the divergence being determined by \mbox{$|\cos\theta_2|^{-1/2}$}. By expanding the sine and cosine terms one finds\footnote{
The limiting behavior~(\ref{singul}) may also be obtained from the calculations by Wolf at al.~\cite{quadrupole}. In that, their angular deflections~$\delta\theta$ and~$\delta\theta''$ correspond to our own~$\theta_1$ and~$\theta_2$ as: \mbox{$\delta\theta=\theta_1$} and \mbox{$\delta\theta''=\pi/2-\theta_2$}. Their equations~(7) and~(9) report only the limiting forms of solid angles from (\ref{solid1}) and (\ref{solid2}):
\begin{align*}
&\Omega_1\to\pi(\delta\theta)^2 \quad\;\;\text{as}\quad \delta\theta\to0,\\
&\Omega_2\to2\pi(\delta\theta'') \quad\text{as}\quad \delta\theta''\to0.
\end{align*}
In the same limit, it follows from~(\ref{solid}) that:
\begin{equation*}
\delta\theta=2\sqrt{\delta\theta''} \qquad\Leftrightarrow\quad \theta_1=2\sqrt{\pi/2-\theta_2}.
\end{equation*}
Applying a derivative from~(\ref{transfor}) immediately yields a limiting behavior from~(\ref{singul}), consisting of an integrable divergence \mbox{$(\pi/2-\theta_2)^{-1/2}$}, pertaining to the upper charge (\mbox{$\theta_2<\pi/2$}).
}:
\begin{equation}
\frac{\varrho_2(\theta_2)}{\varrho_0}\to\frac{1}{\sqrt{|\pi/2-\theta_2|}} \quad\text{as}\quad \theta_2\to\pi/2.
\label{singul}
\end{equation}
This observation confirms that the singularity is indeed integrable, as it should be due to conservation of the density norm from~(\ref{conser}).

\subsection*{3D case}

As addressed in the main paper, Wolf et al.~\cite{quadrupole} have proven that in the divergence-free field regions a density of the field lines in a three-dimensional space \textit{is}, in fact, proportional to a magnitude of an appropriate field component (here an electric field~$\mathbf{E}$). In particular, a local \textit{areal}\footnote{
As a `remnant' of the Gauss's law, a density from~(\ref{Srho_relation}) is an \textit{areal} density---defined per unit area---as opposed to the \textit{angular} densities $\varrho_1$ and $\varrho_2$, defined per unit angle. Striving to avoid a notational clutter, we are using a single notation~$\varrho$ for any kind of field line density. Their meaning and their mutual (in)compatibility should be clear from their definitions and a general context.
} density \mbox{$\varrho(\mathbf{r},\hat{\mathbf{n}})$} of field lines passing through an oriented surface element \mbox{$\D\mathbf{A}=\D A\,\hat{\mathbf{n}}$} is proportional to a field component \textit{perpendicular} to the surface element, i.e. to a component alongside a surface normal~$\hat{\mathbf{n}}$:
\begin{equation}
\varrho(\mathbf{r},\hat{\mathbf{n}})\propto \mathbf{E}(\mathbf{r})\cdot \hat{\mathbf{n}}.
\label{Srho_relation}
\end{equation}
We will demonstrate the validity of this relation on a present linear quadrupole example.

If we wished to do it numerically, by generating a finite set of field lines (just as in \fig~\ref{Sfig3}), we would not even have to play games with an appropriate three-dimensional angular spacing of the field lines. We first consider that a three-dimensional angular density~$\varrho_\text{3D}$ is determined by a solid angle element $\D^2 \Omega$:
\begin{equation}
\varrho_\text{3D}(\theta,\varphi)\equiv\frac{\D^2 N}{\D^2 \Omega}=\frac{\D^2 N}{\sin\theta\,\D\theta\D\varphi},
\label{ang_dens_3d}
\end{equation}
decomposed here in spherical coordinates $\theta$ and $\varphi$. As we are interested only in the polar angle~$\theta$, we will retain a notation~$\varrho$ for an angular density over~$\D\theta$, as in~(\ref{ang_dens}). It is now clear from~(\ref{ang_dens_3d}) that the amount of field lines within an angular opening~$\D\theta$ (and integrated over~$\varphi$) must follow a sine-dependence \mbox{$\D N\propto\sin\theta\,\D\theta$} for a 3D angular distribution to be uniform. This means that a marginal distribution from~(\ref{ang_dens}), when applied to the outgoing field lines around the outer charge, must now be:
\begin{equation}
\varrho_1(\theta_1)=\varrho_0\sin\theta_1
\label{theta_iso}
\end{equation}
in order to represent the field lines isotropically distributed in 3D space. In that, we keep~$\varrho_0$ as a constant parameter.

The notion from~(\ref{theta_iso}) allows us to `recycle' the previously generated field lines ($10^5$ per outer charge) in investigating an initially-3D-isotropic distribution of field lines, without having to generate a new set of field lines following a $\sin\theta_1$ angular dependence. In building a distribution of terminating angles by histogramming them (just as in \fig~\ref{Sfig3}), one only needs to \textit{weight} the contribution of each terminating angle~$\theta_2$ by a factor $\sin\theta_1$. In other words, each $\theta_2$-count is to increase the histogram content not by 1, but rather by~$\sin\theta_1$. Carrying out his procedure, we find a modified distribution of terminating angles to follow a perfect sine-dependence: \mbox{$\varrho_2(\theta_2)/\varrho_0=\sin\theta_2$}, representing a 3D-isotropic distribution of field lines around the central charge. This is a clear signature of the isotropy of an electric field in the immediate vicinity of a point charge, supporting the validity of relation~(\ref{Srho_relation})\footnote{
In a 3D space, both angular densities $\varrho_1$ and $\varrho_2$ obviously correspond to a density of the field lines piercing all surface elements (integrated over~$\varphi$) of the form \mbox{$\D\mathbf{A}=\D A\,\hat{\mathbf{r}}$}, surrounding a given point charge. Hence, in the context of $\varrho_1$ and $\varrho_2$ the relation~(\ref{Srho_relation}) extracts, via \mbox{$\hat{\mathbf{n}}=\hat{\mathbf{r}}$}, a radial field component, which is indeed isotropic in the immediate vicinity of a point charge.
}.

\vspace*{12.5mm}

\pagebreak

We can also exactly demonstrate this result from the previous calculations. We only need to plug~(\ref{theta_iso}) into~(\ref{transfor}) in order to find:
\begin{equation}
\frac{\varrho_2(\theta_2)}{\varrho_0}=\left|\frac{\D\theta_1}{\D\theta_2}\right|\sin\theta_1,
\end{equation}
which is furnished by an additional~$\sin\theta_1$ factor, relative to a 2D-isotropic case. A derivative term we already have from (\ref{quad_analytic}) and we only need to plug $\theta_1$ from (\ref{th_1}) under a sine term, whereupon it remains:
\begin{equation}
\varrho_2(\theta_2)=2\varrho_0\sin\theta_2.
\end{equation}
This is a 3D-isotropy in~$\theta_2$, identical in form to a 3D-isotropy in~$\theta_1$ from~(\ref{theta_iso}). Additional factor of~2 is the same one that we have already encountered in a context of \fig~\ref{Sfig3}.

\section{`Repulsive dipole'}

The next example shows that even a simple configuration of two equal point charges exhibits the effect of equatorial clumping. In the main paper we show this configuration by the right plot from \fig~\ref{fig2} and call it a `repulsive dipole'. Though in certain aspects less elegant than a linear quadrupole, a great advantage of this setup is the availability of the \textit{analytical parametrization} of the field lines. It eliminates the necessity for obtaining them numerically from a differential field line equation. Some numerical calculations will still be necessary, but they will not be raising the typical questions regarding the stability of numerical solutions to a differential equation.

Equations~(\ref{zk_main}) and~(\ref{rk_main}) from the main paper report the field line parametrization for both the `attractive dipole' (original parametrization by Kristjansson~\cite{dipole}) and the `repulsive dipole', both composed of two point charges located at \mbox{$\mathbf{r}_{1,2}=\pm L\,\hat{\mathbf{z}}$}. For simplicity, we use here the dimensionless field line coordinates~$\rho$ and~$z$, redefining those from the main paper as: \mbox{$\rho/L\to\rho$} and \mbox{$z/L\to z$}. Using this convention and focusing on a `repulsive dipole', the relevant parametric field line equations become:
\begin{align}
&z(t)=\pm t\sqrt{t\frac{2-\kappa t}{2t-\kappa}},
\label{zk}\\
&\rho(t)= (t^2-1)\sqrt{\frac{\kappa}{2t-\kappa}}.
\label{rk}
\end{align}
In analytic calculations we will again consider only the field lines above the equator: \mbox{$z>0$}, corresponding to a positive sign in~(\ref{zk}). All of them originate from the upper charge at \mbox{$\mathbf{r}_1= L\,\hat{\mathbf{z}}$}. With that, the factor~$\kappa$---parameterizing a particular field line---is related to the initial field line inclination~$\theta$ (as it leaves the source charge) as \mbox{$\kappa=1+\cos\theta$}. The central running parameter~$t$ spans a range \mbox{$t\in\langle \kappa/2,1]$}.

\begin{figure*}[t!]
\centering
\includegraphics[height=0.3\textwidth,keepaspectratio,angle=0]{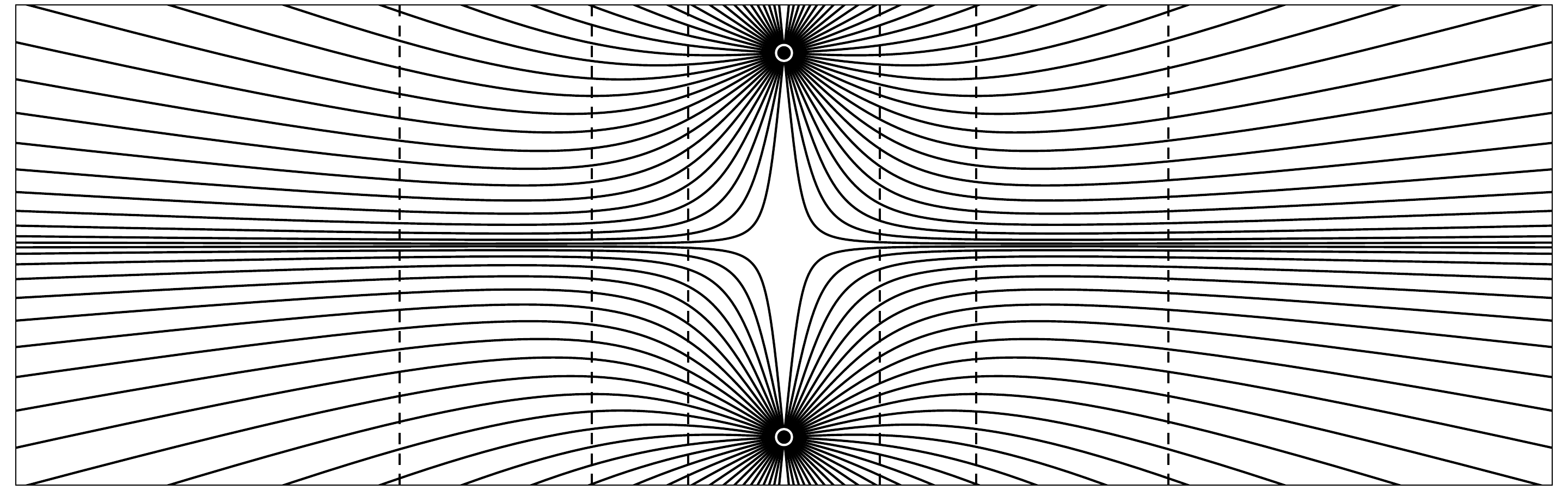}
\caption{Field lines of a `repulsive dipole' configuration consisting of two equal charges, analytically parameterized by~(\ref{zk}) and~(\ref{rk}) in dimensionless coordinates. Vertical dashed lines show three distances from the `dipole' axis ($z$-axis) that we showcase in later \fig~\ref{Sfig5}: \mbox{$\rho=0.5$}, \mbox{$\rho=1$} and \mbox{$\rho=2$}.}
\label{Sfig4}
\vspace*{5mm}
\end{figure*}


\pagebreak

\begin{figure}[b!]
\centering
\includegraphics[width=1\linewidth,keepaspectratio]{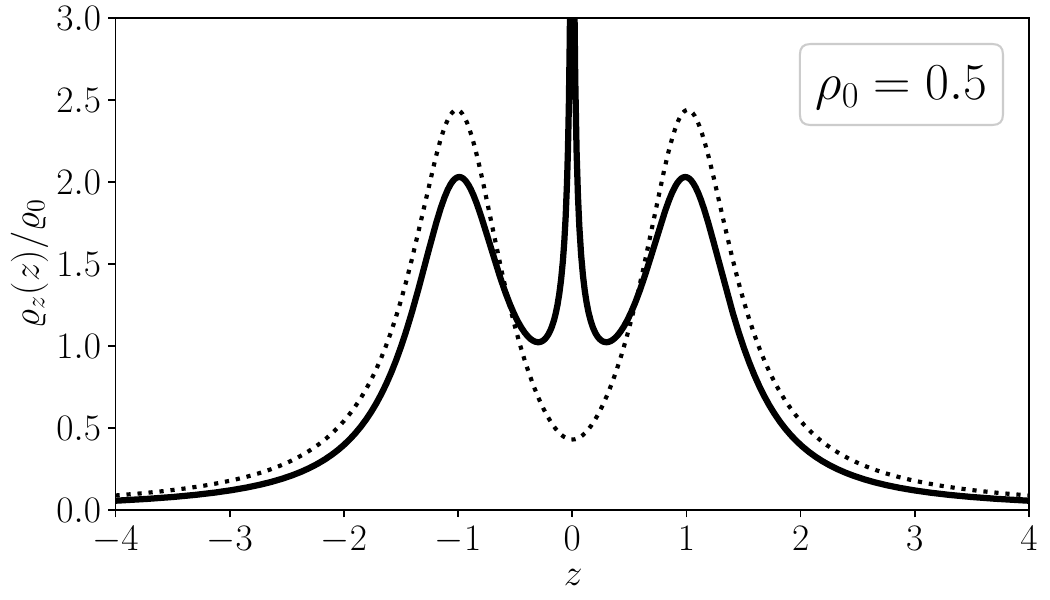}
\includegraphics[width=1\linewidth,keepaspectratio]{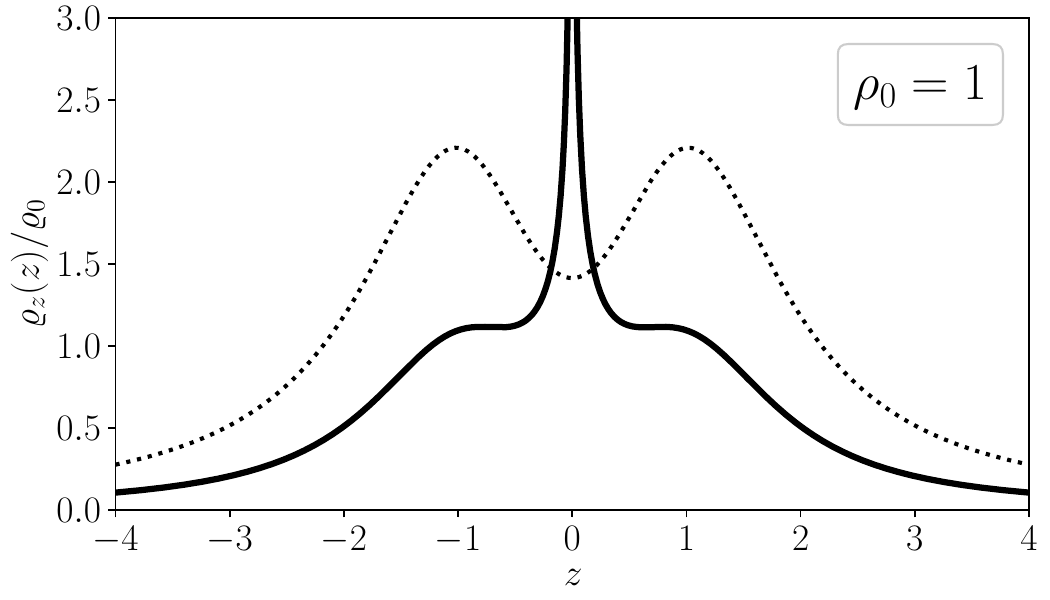}
\includegraphics[width=1\linewidth,keepaspectratio]{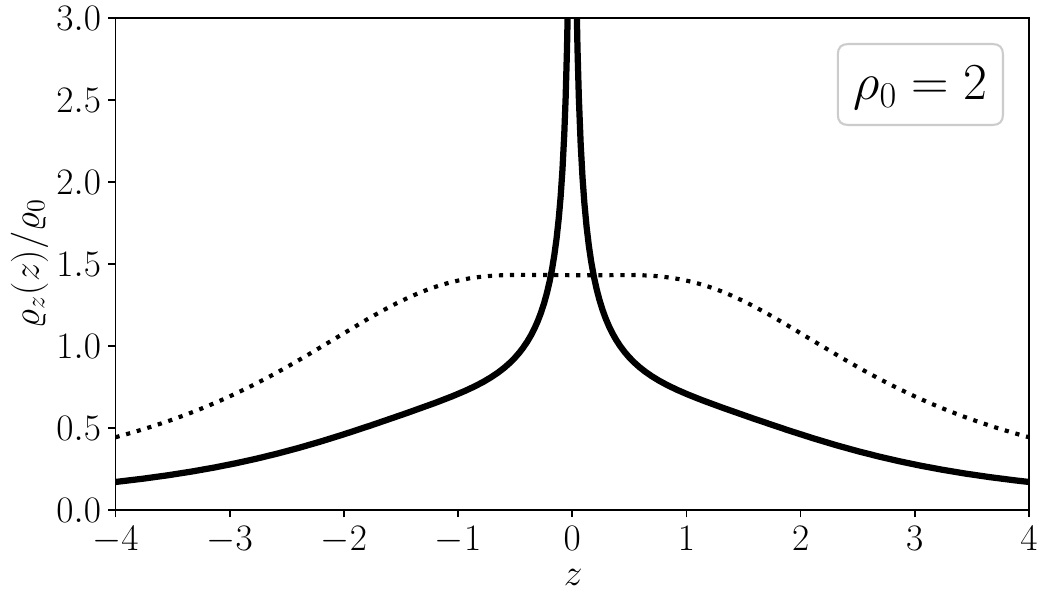}
\caption{Relative field line density for a planar configuration of field lines from \fig~\ref{Sfig4}, at three selected distances from the `dipole' axis. Dotted lines show an arbitrarily scaled magnitude of a total electric field.} 
\label{Sfig5}
\end{figure}

A set of these field lines is shown in \fig~\ref{Sfig4}, emanating in uniform angular steps from the two charges. Vertical dashed lines at \mbox{$\rho=0.5$}, \mbox{$\rho=1$} and \mbox{$\rho=2$} mark the distances from the `dipole' axis that we showcase in later \fig~\ref{Sfig5}. For visual clarity, the vertical field lines lying on the $z$-axis, and the horizontal ones at \mbox{$z=0$} (visible in \fig~\ref{fig2} from the main paper) are not shown here.

The visual suggestion of equatorial clumping is once more perfectly clear. And once again we may speculate that the density of this initially-2D-isotropic field line configuration diverges toward the \mbox{$z=0$} axis. This is indeed the case, as we will soon show. Let us first put forth the `rules of the game'. As with the linear quadrupole, we introduce an \textit{angular} density~$\varrho_\theta$ of the field lines emanating from the upper charge at~$\mathbf{r}_1$:
\begin{equation}
\varrho_\theta(\theta)\equiv\frac{\D N}{\D \theta} \quad\text{as}\quad \mathbf{f}\to\mathbf{r}_1,
\end{equation}
wherein \mbox{$\mathbf{f}=\rho(t)\hat{\boldsymbol{\rho}}+z(t)\hat{\mathbf{z}}$} stands for a vectorial field line parametrization by~(\ref{zk}) and (\ref{rk}). Following the usual rules of the planar field line diagrams, we require that this initial density be uniform ($\varrho_0=\text{const.}$):
\begin{equation}
\varrho_\theta(\theta)=\varrho_0,
\label{unifor}
\end{equation}
as it represents the isotropy of an electrostatic field in the immediate vicinity of a point charge. We now wish to investigate a \textit{linear} density~$\varrho_z$ of the field lines at some fixed distance $\rho_0$ from the `dipole' axis:
\begin{equation}
\varrho_z(z)\equiv\frac{\D N}{\D z} \quad\text{for}\quad \rho=\rho_0.
\end{equation}
Though~$\varrho_z(z)$, together with later quantities, is also dependent on~$\rho_0$, we will suppress any explicit designation of it and consider this dependence implicitly understood.

As with a linear quadrupole, we could now build a spectrum of~$\varrho_z$ by considering a multitude of field lines and `counting'  their $z$-coordinates at a given~$\rho_0$ (e.g. along some of the dashed lines from \fig~\ref{Sfig4}). But this time,  instead of having to obtain each particular field line by a numerical solution to a differential field line equation, we can take advantage of the available analytical parametrization. For a given~$\rho_0$ and a selection of uniformly distributed initial field line inclinations~$\theta$, we first need to obtain (for each $\theta$-defined field line) a value~$t_\theta$ of the parameter~$t$ from~(\ref{rk}), defining a point at a distance~$\rho_0$ from the `dipole' axis. Explicitly writing out~$\kappa$ and rearranging the equation, this boils down to solving:
\begin{equation}
\rho_0^2(2t_\theta-1-\cos\theta)-(1+\cos\theta)(1-t_\theta^2)^2=0
\label{t_theta}
\end{equation}
for~$t_\theta$. Evidently, this is a fourth-degree polynomial in~$t_\theta$. Its closed-form solution does exist, but it is unwieldy so the equation is best solved numerically. After that the sought $z_\theta$ values to be `counted' may be calculated from~(\ref{zk}):
\begin{equation}
z_\theta= t_\theta\sqrt{t_\theta\frac{2-(1+\cos\theta) t_\theta}{2t_\theta-1-\cos\theta}}.
\label{z_theta}
\end{equation}
And that is all there is to it.\\

Still, we will lay out the rest of the (numerical) procedure for calculating~$\varrho_z$ at any given point, without relying on a high-density counting. An analytical detour will also allow us to demonstrate a divergence in~$\varrho_z$. From the same `field lines conservation' considerations as in~(\ref{conser})---only now of the form $\varrho_\theta(\theta)|\D\theta|=\varrho_z(z)|\D z|$---implementing~(\ref{unifor}) we obtain:
\begin{equation}
\frac{\varrho_z(z)}{\varrho_0}=\left|\frac{\D\theta}{\D z}\right|.
\label{transfor_zth}
\end{equation}
As we wish to have~$\varrho_z$ as a function of~$z$, we first need to numerically solve~(\ref{z_theta}) for~$\theta_z$: an initial inclination of a particular field line that passes through the coordinates \mbox{$(\rho_0,z)$}. Setting up a numerical procedure for~$\theta_z$, thus obtained values may easily be numerically differentiated so as to obtain a sought linear density~(\ref{transfor_zth}), yielding the same result a proper `counting' of~$z_\theta$ values would. 

\pagebreak

In summary, this course of action comprises the following three numerical procedures:
\begin{itemize}\itemsep0em
\item[(1)] implementation of a numerical search for \mbox{$t_\theta\in\langle (1+\cos\theta)/2,1]$} by solving~(\ref{t_theta}) for inputs \mbox{$\theta\in\langle0,\pi\rangle$} and \mbox{$\rho_0>0$};
\item[(2)] implementation of a numerical search for~$\theta_z\in[0,\pi]$ by solving~(\ref{z_theta}) for inputs \mbox{$z>0$} and \mbox{$\rho_0>0$}, incorporating a previous implementation for~$t_\theta$;
\item[(3)] numerical differentiation of~$\theta_z$, as per~(\ref{transfor_zth}), for inputs \mbox{$z>0$} and \mbox{$\rho_0>0$}.
\end{itemize}
All these procedures are readily available in modern programming languages and are remarkably stable, robust and dependable.


\pagebreak


An alternative route is also available, sidestepping numerical differentiation at the price of somewhat more involved algebraic expressions. It may be of some interest so we report it here. As we can not hope to obtain a closed-form expression for $\theta_z$ from~(\ref{z_theta}), we express a derivative from~(\ref{transfor_zth}) as:
\begin{equation}
\frac{\D\theta_z}{\D z}\bigg|_z=\bigg(\frac{\D z_\theta}{\D\theta}\bigg|_{\theta_z}\bigg)^{-1}.
\label{fancy}
\end{equation}
The required derivative \mbox{$\D z_\theta/\D\theta$} may now be manually calculated from~(\ref{z_theta}), allowing for the appearance of as yet unresolved derivative \mbox{$\D t_\theta/\D\theta$}. To this end~$z_\theta$ is best expressed as \mbox{$z_\theta=t_\theta^{3/2}\sqrt{\chi}$}, with~$\chi$ denoting a fractional term below a square root in~(\ref{z_theta}). Some algebra yields:
\begin{equation}
\frac{\D z_\theta}{\D\theta}=\frac{3z_\theta}{2t_\theta}\frac{\D t_{\theta}}{\D \theta}+\frac{t_\theta^3}{2z_\theta}\frac{\D\chi}{\D\theta}
\label{dz_dth}
\end{equation}
with:
\begin{equation}
\frac{\D\chi }{\D\theta}=\frac{2(t_\theta^2-1)\sin\theta+(\cos\theta+3)(\cos\theta-1)\frac{\D t_{\theta}}{\D \theta}}{(2t_{\theta}-1-\cos\theta)^2}.
\label{dch_dth}
\end{equation}
Finally, the missing derivative \mbox{$\D t_\theta/\D\theta$} is obtained by differentiating~(\ref{t_theta}), whereupon we are left with:
\begin{equation}
\frac{\D t_\theta}{\D \theta}=-\frac{\rho_0^2+(1-t_\theta^2)^2}{2\rho_0^2+4(1+\cos\theta)(1-t_\theta^2)t_\theta}\sin\theta.
\label{dt_dth}
\end{equation}
Complementing~(\ref{dz_dth}) with~(\ref{z_theta}), (\ref{dch_dth}) and~(\ref{dt_dth}), the task of obtaining the derivative \mbox{$\D z_\theta/\D\theta$} is entirely reduced to finding a numerical solution~$t_\theta$. As per~(\ref{fancy}), an additional numerical search for~$\theta_z$ must be performed, so as to evaluate a derivative at a value of~$\theta$ corresponding to a desired value of~$z$.


To summarize, the present procedure, in its entirety, comprises the following numerical steps:
\begin{itemize}\itemsep0em
\item[(1)] implementation of a numerical search for \mbox{$t_\theta\in\langle (1+\cos\theta)/2,1]$} by solving~(\ref{t_theta}) for inputs \mbox{$\theta\in\langle0,\pi\rangle$} and \mbox{$\rho_0>0$};
\item[(2)] a numerical search for~$\theta_z\in[0,\pi]$ by solving~(\ref{z_theta}) for inputs \mbox{$z>0$} and \mbox{$\rho_0>0$}, incorporating a previous implementation for~$t_\theta$;
\item[(3)] numerical search for~$t_\theta$, based on the previous value of~$\theta_z$, to be used in a direct evaluation of a derivative~(\ref{dz_dth}).
\end{itemize}
Evidently, the first two steps are basically the same as before, while the numerical differentiation is replaced by another algebraic search for~$t_\theta$.\\

Whichever route is taken---$z_\theta$ `counting', a numerical procedure involving numerical differentiation or the one sidestepping it---they all consistently lead to the same results. We show some examples in \fig~\ref{Sfig5}. Separate panels show the relative field line density from~(\ref{transfor_zth}) at three distances from the `dipole' axis, previously marked in \fig~\ref{Sfig4}. For comparison, dotted lines show the arbitrarily scaled magnitude of an electric field. As in \fig~\ref{Sfig4}, the central field line density divergence strongly suggests itself.

\pagebreak


We will now demonstrate that a divergence is indeed present. To this end we investigate a limiting density behavior as \mbox{$z\to0$}, to leading order in some small parameter. Since the field lines approaching the \mbox{$z=0$} axis are those that are from the upper charge emanated `downwards', at angles \mbox{$\theta\approx\pi$}, we introduce:
\begin{equation}
x\equiv\pi-\theta,
\end{equation}
which will serve as a small positive expansion parameter.

We must first identify a limiting behavior of~$t_\theta$ from~(\ref{t_theta}). Let us observe a limiting behavior of the entire equation as \mbox{$\theta\to\pi$}. From \mbox{$1+\cos\theta\to0$} it follows:
\begin{equation}
2\rho_0^2t\to0 \quad\text{as}\quad \theta\to\pi.
\label{limit_cond}
\end{equation}
As \fig~\ref{Sfig4} suggests and~(\ref{rk}) confirms, for any \mbox{$\theta\neq0,\pi$} (we being interested in~$\theta$ just below~$\pi$) the field lines span the entire range \mbox{$\rho\in[0,\infty\rangle$}. This means that the limiting form~(\ref{limit_cond}) may produce nontrivial values \mbox{$\rho_0>0$} only if:
\begin{equation}
t_\theta\to0 \quad\text{as}\quad \theta\to\pi.
\label{t_to0}
\end{equation}
For small~$t_\theta$ we may assume, to leading order in~$x$, a limiting behavior \mbox{$t_\theta\to ax^n+b$}; $n$ being as yet unspecified lowest power dominating the dependence on~$x$. Since~$t_\theta$ must not be divergent as \mbox{$x\to0$}, we have a constraint \mbox{$n>0$}. On the other hand, from~(\ref{t_to0}) we immediately have \mbox{$b=0$}. Therefore, we are left with the simple form \mbox{$t_\theta\to ax^n$} to be introduced into the limiting behavior of~(\ref{t_theta}). In order to determine~$a$ and~$n$ we expand the $\cos\theta$ term from~(\ref{t_theta}) around \mbox{$\theta=\pi$}:
\begin{equation}
\cos\theta\to-1+\tfrac{(\pi-\theta)^2}{2}=-1+\tfrac{x^2}{2} \quad\text{as}\quad x\to0.
\end{equation}
Introducing that into~(\ref{t_theta}) we obtain a limiting behavior:
\begin{equation}
(1+\rho_0^2)x^2-4a\rho_0^2 x^n-2a^2x^{2+2n}+a^4x^{2+4n}=0 \quad\text{as}\quad x\to0.
\label{cond_exp}
\end{equation}
Since \mbox{$n>0$}, the terms $x^{2+2n}$ and $x^{2+4n}$ immediately have an exponent strictly greater than~2. Due to the presence of term $x^2$ they drop out of the limiting behavior. After that only the terms $x^2$ and $x^n$ remain to produce a zero from the right hand side of equation. This is only possible with~\text{$n=2$}. Plugging this back into~(\ref{cond_exp}), while keeping only the dominant terms, $a$ immediately follows:
\begin{equation}
(1+\rho_0^2-4a\rho_0^2)x^2=0 \quad\Rightarrow\quad a=\tfrac{1}{4}(1+\rho_0^{-2}).
\end{equation}
With that we have identified a limiting behavior:
\begin{equation}
t_\theta\to \tfrac{1+\rho_0^{-2}}{4}\,x^2 \quad\text{as}\quad x\to0.
\end{equation}
Plugging this into~(\ref{z_theta}) we obtain to leading order in~$x$:
\begin{equation}
z_\theta\to\tfrac{(1+\rho_0^2)^{3/2}}{4\rho_0^2} \,x^2 \quad\text{as}\quad x\to0.
\end{equation}
From this we only need to extract an inverse dependence:
\begin{equation}
\theta_z\to\pi-\tfrac{2\rho_0}{(1+\rho_0^2)^{3/4}}\sqrt{z} \quad\text{as}\quad z\to0
\label{theta_limit}
\vspace*{1mm}
\end{equation}
in oder to obtain from~(\ref{transfor_zth}):
\begin{equation}
\frac{\varrho_z(z)}{\varrho_0}\to\frac{\rho_0}{(1+\rho_0^2)^{3/4}} \, \frac{1}{\sqrt{|z|}} \quad\text{as}\quad z\to0,
\label{divergence}
\end{equation}
where we have immediately generalized the result for \mbox{$z<0$}. Therefore, a field line density is indeed divergent, and the divergence is integrable, as it should be.


\vspace*{-1mm}

\subsection*{3D case}

We confirm that the initially-3D-isotropic distribution $\varrho_\theta(\theta)=\varrho_0\sin\theta$ of field lines, as they leave the upper charge, produces a relative density:
\begin{equation}
\frac{\varrho_z(z)}{\varrho_0}=\left|\frac{\D\theta}{\D z}\right|\sin\theta
\label{varrho_3d}
\end{equation}
which, at a given~$\rho_0$, follows a $z$-dependence of a field component~$E_\rho$ (expressed in dimensionless coordinates):
\begin{equation}
E_\rho(z;\rho_0)\propto \frac{\rho_0}{[(z-1)^2+\rho_0^2]^{3/2}}+\frac{\rho_0}{[(z+1)^2+\rho_0^2]^{3/2}},
\label{E_rho}
\end{equation}
in accordance with a general result~(\ref{Srho_relation}). In fact, we can demonstrate that the sine term from~(\ref{varrho_3d}) does indeed remove a divergence from the \mbox{$\D\theta/\D z$} term. To this end we only need to insert a limiting behavior of~$\theta_z$ from~(\ref{theta_limit}) in order to find (after generalizing for \mbox{$z<0$}):
\begin{equation}
\sin\theta_z\to\tfrac{2\rho_0}{(1+\rho_0^2)^{3/4}}\sqrt{|z|} \quad\text{as}\quad z\to0,
\end{equation}
to leading order in~$z$. Combining that with~(\ref{divergence}), corresponding to a limiting behavior of \mbox{$|\D\theta/\D z|$}, yields:
\begin{equation}
\frac{\varrho_z(z)}{\varrho_0}\to\frac{2\rho_0^2}{(1+\rho_0^2)^{3/2}} \quad\text{as}\quad z\to0.
\end{equation}
A finite value, bearing a clear resemblance to what remains of a $z$-dependence from~(\ref{E_rho}) for \mbox{$z=0$}.


%

\onecolumngrid

\end{document}